\documentclass[12pt]{article}

\def\eptwo{\left\{ \phantom{|}^{\mu\nu}_{ab} \right\}}
\def\epthree{\left\{ \phantom{|}^{\mu\nu\alpha}_{abc} \right\}}
\def\epfour{\left\{ \phantom{|}^{\mu\nu\alpha\beta}_{abcd} \right\}}
\def\epfive{\left\{ \phantom{|}^{\mu\nu\alpha\beta\gamma}_{abcde}%
\right\}}

\usepackage{color}
\usepackage{pict2e}

\author{M.~V.~Khabarov${}^{a,b}$\thanks{maksim.khabarov@ihep.ru},
Yu.~M.~Zinoviev${}^{b,a}$\thanks{yurii.zinoviev@ihep.ru}
\\[0.5cm]
\it\small{${}^a$Moscow Institute of Physics and Technology (State
University),} \\
\it{\small Dolgoprudny, Moscow Region, 141701, Russia} \\[0.3cm]
\it{\small ${}^b$Institute for High Energy Physics of National
Research Center "Kurchatov Institute"}\\
\it{\small Protvino, Moscow Region, 142281, Russia}}

\title{Infinite (continuous) spin fields \\
in the frame-like formalism}

\date{}

\begin{document}

\maketitle

\begin{abstract}
In this paper we elaborate on the gauge invariant frame-like
Lagrangian description for the wide class of the so-called infinite
(or continuous) spin representations of Poincar\'e group. We use our
previous results on the gauge invariant formalism for the massive
mixed symmetry fields corresponding to the Young tableau with two rows
$Y(k,l)$ ($Y(k+1/2,l+1/2)$ for the fermionic case). We have shown that
the corresponding infinite spin solutions can be constructed as a
limit where $k$ goes to infinity, while $l$ remain to be fixed and
label different representations. Moreover, our gauge invariant
formalism provides a natural generalization to (Anti) de Sitter
spaces as well. As in the completely symmetric case considered earlier
by Metsaev we have found that there are no unitary solutions in
de Sitter space, while there exists a rather wide spectrum of Anti
de Sitter ones. In this, the question what representations of the Anti
de Sitter group such solutions correspond to remains to be open.
\end{abstract}

\thispagestyle{empty}
\newpage
\setcounter{page}{1}

\tableofcontents

\section*{Introduction}

Besides the very well known finite component massless and massive
representations of the Poincar\'e algebra there exists a number of
rather exotic so-called infinite (or continuous) spin ones (see e.g.
\cite{BKRX02,BB06}). In dimensions $d \ge 4$ they have an infinite
number of physical degrees of freedom and so may be of some interest
for the higher spin theory. Indeed, they attracted some attention last
times \cite{BM05,Ben13,ST14,ST14a,Riv14,BNS15} (see also recent review
\cite{BS17}). One of the reason is that such representation being
massless nevertheless are characterized by some dimensionful parameter
$\mu$, which may play important role in the possibility to construct
interactions with such fields.

It has been already noted several times in different contexts that
such infinite spin representations may be considered as a limit of
massive higher spin ones where spin goes to infinity, mass goes to
zero while their product $ms = \mu$ remains to be fixed and provides
this dimensionful parameter that characterizes the representation.
A very nice Lagrangian realization for this idea has been given
recently by Metsaev in \cite{Met16} for the bosonic case and in
\cite{Met17} for the fermionic one (see also \cite{MN17}). Namely, he
has shown that the very same gauge invariant formalism that was
previously used for the description of massive higher spin bosonic
\cite{Zin01} and fermionic \cite{Met06} particles can be used to
provide Lagrangian gauge invariant formulation for the infinite spin
cases. It is interesting that even in flat Minkowski space there
exist unitary models that resemble the so-called partially massless
models for the finite spin particles in de Sitter space. Recall
that one of the nice features of such gauge invariant formulation for
the massive higher spins is that it perfectly works not only in flat
Minkowski space but in (Anti) de Sitter spaces as well, giving a
possibility to investigate all possible massless and partially
massless limits. The same holds for the infinite spin cases as well
and it appeared that while there are no unitary infinite spin
models in de Sitter space there exists quite a lot of different
unitary solutions in Anti de Sitter space though till now it is
not clear what representations of the Anti de Sitter group they
correspond to.

Naturally, the most important open physical question is the
possibility to have consistent interacting theories containing such
infinite spin fields. A very important step in this direction was
recently made by Metsaev \cite{met17a}, who using light-cone formalism
provided a classification of cubic interaction vertices for one
massless infinite spin field with two massive finite spin ones as
well as for two massless infinite spin fields with one massive finite
spin one (see also \cite{BMN17} for the interaction of two massive
scalars with one massless infinite spin field).

In general, the classification of the infinite spin representations is
provided by the representations of the so-called short little group
$SO(d-3)$ \cite{BKRX02,BB06}. It is clear that in $d=3$ and $d=4$
dimensions this short little group is trivial so we have just one
bosonic and one fermionic infinite spin representation whose
description is based on the completely symmetric (spin-)tensors. But
in dimensions higher than 4 we face a huge number of such
representations. For example, in $d=5,6$ such representations are
labeled by the parameter $l$ that can take integer or half-interger
values for the bosonic and fermionic cases respectively. It seemed
natural to assume that the Lagrangian formalism for such
representations can be obtained starting with the gauge invariant
description for the massive mixed symmetry fields corresponding to the
Young tableau with two rows $Y(k,l)$ ($Y(k+1/2,l+1/2)$ for the
fermionic case) if one takes a limit where $k$ goes to infinity, while
$l$ being fixed and labels different representations. In this paper
using our previous results \cite{Zin08b,Zin08c,Zin09b,Zin09c} we show
that it appears to be the case. As in the completely symmetric case
our gauge invariant formalism provides an extension to (Anti) de
Sitter space and we also found that there are no unitary solutions
in de Sitter space while there exists a number of unitary ones in
Anti de Sitter space. Let us stress ones again that what
representations of the Anti de Sitter group such solutions correspond
to is still an open question.

Our paper is organized as follows. We work in the gauge invariant
frame-like formalism\footnote{See \cite{Zin17} for the
three-dimensional case}. In Section 1 we provide all necessary
information  for the frame-like description of the massless finite
component mixed symmetry bosonic fields that serve as the natural
building blocks for the infinite spin ones. In section 2 we begin with
the completely symmetric case providing frame-like generalization of
Metsaev's results \cite{Met16}. Then we consider a simplest mixed
symmetry bosonic example based on the so-called long hooks $Y(k,1)$
and then we elaborate on the general mixed symmetry case $Y(k,l)$, $k
\to \infty$. Similarly, in Section 3 we provide all necessary
information  for the frame-like description of the massless finite
component mixed symmetry fermionic fields while Section 4  contains
three subsections analogous to the ones in Section 2.

{\bf Notations and conventions} We work in the frame-like
formalism where  the world indices are denoted by the Greek letters,
while the local indices --- by the Latin ones. To simplify formulas we
use the so-called condensed notations for the local indices so
that
$$
\Phi^{a(k),b(l)} = \Phi^{(a_1a_2\dots a_k),(b_1b_2\dots b_l)}
$$
where round brackets denote symmetrization which uses the minimum
number of terms necessary without any normalization factor. For the
local indices denoted by the same letter and placed on the same level
we always assume symmetrization, e.g.
$$
e^a \Phi^{a(k),b(l)} = e^{(a_1} \Phi^{a_2\dots a_{k+1}),b(l)}
$$
Working with the fermions we heavily use the completely
antisymmetric products of $\gamma$-matrices defined as follows
$$
\Gamma^{a[n]} = \frac{1}{n!} \gamma^{[a} \dots \gamma^{a]}
$$
with a couple of useful relations
\begin{eqnarray*}
\gamma^a \Gamma^{b[n]} &=& g^{ab} \Gamma^{b[n-1]} + \Gamma^{ab[n]} \\
\Gamma^{a[n]b} \gamma_b &=& (d-n) \Gamma^{a[n]}
\end{eqnarray*}
(Anti) de Sitter space is described by the (non-dynamic)
background frame $e_\mu{}^a$ and the covariant derivative $D_\mu$. All
our expressions for the Lagrangians and gauge transformations are
completely antisymmetric on the world indices so the covariant
derivatives always act on the local indices only (including an
implicit spinor index). For the bosonic and fermionic objects the
commutator of the covariant derivatives are normalized as follows
\begin{eqnarray*}
\ [ D_\mu, D_\nu ] \xi^{a(k)} &=& - \kappa e_{[\mu}{}^a
\xi_{\nu]}{}^{a(k-1)}, \qquad \kappa = \frac{2\Lambda}{(d-1)(d-2)} \\
\ [ D_\mu, D_\nu ] \zeta^{a(k)} &=& - \kappa [ e_{[\mu}{}^a
\zeta_{\nu]}{}^{a(k-1)} + \frac{1}{2} \Gamma_{\mu\nu} \zeta^{a(k)} ]
\end{eqnarray*}
In what follows we also use the completely antisymmetric products
of the inverse frames:
$$
\eptwo = e^{[\mu}{}_a e^{\nu]}{}_b, \qquad
\epthree = e^{[\mu}{}_a e^\nu{}_b e^{\alpha]}{}_c 
$$
and so on.

\section{Massless bosonic fields}

In this section we provide all necessary information on the massless
(finite component) symmetric and mixed symmetry bosonic fields which
will serve as the building blocks for our construction of the
infinite component cases. In what follows $Y(k,l)$ denote mixed
symmetry tensor having two rows with length $k$ and $l$ respectively.

\subsection{Completely symmetric tensor $Y(k+1,0)$}

Frame-like description for the completely symmetric tensor
\cite{Vas80,LV88} requires a physical one-form $h_\mu{}^{a(k)}$,
symmetric and traceless on its local indices and an auxiliary one-form
$\omega_\mu{}^{a(k),b}$, symmetric on its first $k$ local indices,
traceless on all local indices and satisfying $\omega_\mu{}^{a(k),a} =
0$. In flat Minkowski space the free Lagrangian looks like:
\begin{eqnarray}
(-1)^k {\cal L}_0(h_\mu{}^{a(k)}) &=& - \eptwo [ 
\omega_\mu{}^{ae(k-1),c} \omega_\nu{}^{be(k-1),c} + \frac{1}{k}
\omega_\mu{}^{e(k),a} \omega_\mu{}^{e(k),b} ] \nonumber \\
 && - 2 \epthree \omega_\mu{}^{ae(k-1),b} \partial_\nu
h_\alpha{}^{ce(k-1)} \label{lagb1}
\end{eqnarray}
This Lagrangian is invariant under the following gauge
transformations:
\begin{equation}
\delta_0 \omega_\mu{}^{a(k),b} = \partial_\mu \chi^{a(k),b}, \qquad
\delta_0 h_\mu{}^{a(k)} = \partial_\mu \zeta^{a(k)} + 
\chi^{a(k),}{}_\mu \label{gaugeb1}
\end{equation}
where parameters $\chi^{a(k),b}$ and $\zeta^{a(k)}$ have the same
properties on their local indices as the gauge fields 
$\omega_\mu{}^{a(k),b}$ and $h_\mu{}^{a(k)}$ respectively.

After the replacement of the ordinary partial derivatives by the $AdS$
covariant ones the Lagrangian ceases to be invariant:
$$
\delta_0 {\cal L}_0 = (-1)^k \frac{2(k+1)(d+k-3)}{k} \kappa
[ \omega_\mu{}^{a(k),\mu} \zeta^{a(k)} - \chi^{a(k),\mu} 
h_\mu{}^{a(k)} ]
$$
This non-invariance can be compensated with the introduction of the
mass-like terms (recall that in $(A)dS$ space the presence of such
terms does not necessarily means that the field is massive):
\begin{equation}
(-1)^k {\cal L}_2(h_\mu{}^{a(k)}) = b_k \eptwo h_\mu{}^{ae(k-1)} 
h_\nu{}^{be(k-1)} \label{lagb2}
\end{equation}
and corresponding corrections to the gauge transformations:
\begin{eqnarray}
\delta_2 \omega_\mu{}^{a(k),b} &=& \frac{b_k}{(k+1)(d-2)} [ k 
e_\mu{}^b \zeta^{a(k)} - e_\mu{}^a \zeta^{a(k-1)b} \nonumber \\
 && \qquad \qquad - \frac{1}{(d+k-3)} ( (k-1) g^{ba} 
\zeta_\mu{}^{a(k-1)} - 2 g^{a(2)} \zeta_\mu{}^{a(k-2)b}) ]
\label{gaugeb2}
\end{eqnarray}
provided
$$
b_k = (k+1)(d+k-3)\kappa
$$

\subsection{Mixed symmetry tensor $Y(k+1,1)$}

For the frame-like description of the massless mixed symmetry fields
in the flat Minkowski space we follow
\cite{Zin03,Skv08,Skv08a}\footnote{For the description of mixed
symmetry fields in $AdS$ space see e.g. \cite{ASV03,ASV05,ASV06}}.

The frame-like description for the mixed symmetry tensor $Y(k+1,1)$
(so-called long hook) appears to be special and deserves separate
consideration. For this case we use a physical two-form
$\Phi_{\mu\nu}{}^{a(k)}$, completely symmetric and traceless on its
local indices and an auxiliary one-form $\Omega_\mu{}^{a(k),bc}$, with
local indices corresponding to the Young tableau $Y(k,1,1)$, i.e.
symmetric on its first $k$ indices, antisymmetric on the last two
ones, traceless on all indices and satisfying the relation
$\Omega_\mu{}^{a(k),ab} = 0$. In the flat case the free Lagrangian has
the form:
\begin{eqnarray}
(-1)^k {\cal L}_0 &=& \eptwo [ \Omega_\mu{}^{ae(k-1),cd}
\Omega_\nu{}^{be(k-1),cd} + \frac{2}{k} \Omega_\mu{}^{e(k),ac}
\Omega_\nu{}^{e(k),bc} ] \nonumber \\
 && - \epfour \Omega_\mu{}^{ae(k-1),bc} \partial_\nu
\Phi_{\alpha\beta}{}^{de(k-1)} \label{lagb3}
\end{eqnarray}
This Lagrangian is invariant under the following gauge
transformations:
\begin{equation}
\delta_0 \Omega_\mu{}^{a(k),bc} = \partial_\mu \eta^{a(k),bc}, \qquad
\delta_0 \Phi_{\mu\nu}{}^{a(k)} = \partial_{[\mu} \xi_{\nu]}{}^{a(k),}
+ \eta^{a(k)}{}_{\mu\nu} \label{gaugeb3}
\end{equation}
where the gauge parameters $\eta^{a(k),bc}$ and $\xi_\mu{}^{a(k)}$
have the same properties on their local indices as the gauge fields
$\Omega_\mu{}^{a(k),bc}$ and $\Phi_{\mu\nu}{}^{a(k)}$.

The replacement of the ordinary partial derivatives by the $AdS$
covariant ones spoils the invariance of the Lagrangian:
\begin{equation}
(-1)^k \delta {\cal L}_0 = \frac{(k+2)(d+k-4)}{k} \kappa \eptwo [
\eta_\mu{}^{e(k),ab} \Phi_{\mu\nu}{}^{e(k)} - 2 \Omega_\mu{}^{e(k),ab}
\xi_\nu{}^{e(k)} ]
\end{equation}
In this special case there are no mass-like terms that can be added to
the Lagrangian and the invariance of the Lagrangian can not be
restored without introduction of some additional fields (see the next
section). The reason is the essential difference in the spectrum of
the mixed symmetry representations of the Poincar\'e and the $(A)dS$
groups (see \cite{BMV00}).

\subsection{General mixed symmetry tensor $Y(k+1, m+1)$}

In this case we need \cite{Skv08} the two-forms 
$\Psi_{\mu\nu}{}^{a(k),b(m)}$ and $\Omega_{\mu\nu}{}^{a(k),b(m),c}$
with their local indices corresponding to the Young tableau $Y(k,m)$
and $Y(k,m,1)$. The free Lagrangian in the flat case looks like:
\begin{eqnarray}
(-1)^{k+m} {\cal L}_0 &=& - \frac{1}{2} \epfour [ 
\Omega_{\mu\nu}{}^{ae(k-1),b(f(m-1),g} 
\Omega_{\alpha\beta}{}^{ce(k-1),d(f(m-1),g} \nonumber \\
 && \qquad \qquad + \frac{1}{k} \Omega_{\mu\nu}{}^{e(k),a(f(m-1),b}
\Omega_{\alpha\beta}{}^{e(k),c(f(m-1),d} \nonumber \\
 && \qquad \qquad + \frac{1}{m} \Omega_{\mu\nu}{}^{a(e(k-1),f(m),b}
\Omega_{\alpha\beta}{}^{ce(k-1),f(m),d} ] \nonumber \\
 && + \epfive \Omega_{\mu\nu}{}^{af(k-1),bg(m-1),c}
\partial_\alpha \Psi_{\beta\gamma}{}^{df(k-1),eg(m-1)} \label{lagb4}
\end{eqnarray}
It is invariant under the following gauge transformations:
\begin{eqnarray}
\delta_0 \Omega_{\mu\nu}{}^{a(k),b(m),c} &=& \partial_{[\mu}
\eta_{\nu]}{}^{a(k),b(m),c}, \nonumber \\
\delta_0 \Psi_{\mu\nu}{}^{a(k),b(m)} &=& \partial_{[\mu}
\xi_{\nu]}{}^{a(k),b(m)} + \eta_{[\mu}{}^{a(k),b(m),}{}_{\nu]}
\label{gaugeb4}
\end{eqnarray}
where the one-form gauge parameters $\eta_\mu{}^{a(k),b(m),c}$ and 
$\xi_\mu{}^{a(k),b(m)}$ have the same properties on their local
indices as the corresponding gauge fields.

With the replacement of the ordinary partial derivatives by the $AdS$
covariant ones we obtain:
\begin{eqnarray}
(-1)^{k+m} \delta {\cal L}_0 &=& 2\kappa \epthree [ 
\frac{(k+2)(d+k-4) + (d+m-6)}{k} \Omega_{\mu\nu}{}^{d(k),ae(m-1),b}
\xi_\alpha{}^{d(k),c(e(m-1)} \nonumber \\
 && \qquad \qquad + \frac{(m+1)(d+m-6)}{m} 
\Omega_{\mu\nu}{}^{ad(k-1),e(m),b} \xi_\alpha{}^{cd(k-1),e(m)}
\nonumber \\
 && \qquad \qquad + \frac{(k+2)(d+k-4) + (d+m-6)}{k} 
\eta_\mu{}^{d(k),ae(m-1),b} \Psi_{\nu\alpha}{}^{d(k),ce(m-1)}
\nonumber \\
 && \qquad \qquad + \frac{(m+1)(d+m-6)}{m} \eta_\mu{}^{ad(k-1),e(m),b}
\Psi_{\nu\alpha}{}^{cd(k-1),e(m)} ]
\end{eqnarray}
One can try to compensate for this non-invariance by introducing the
mass-like terms
\begin{equation}
(-1)^{k+m} {\cal L}_2 = a_{k,m} \epfour 
\Psi_{\mu\nu}{}^{ae(k-1),bf(m-1)} 
\Psi_{\alpha\beta}{}^{ce(k-1),df(m-1)} \label{lagb5}
\end{equation}
as well as the corresponding corrections to the gauge transformations:
\begin{eqnarray}
\delta_2 \Omega_{\mu\nu}{}^{a(k),b(m),c} &=& - 
\frac{2a_{k,m}}{(k+2)(m+1)(d-4)} [ (k+1)m e_{[\mu}{}^c 
\xi_{\nu]}{}^{a(k),b(m)} - m e_{[\mu}{}^{a} 
\xi_{\nu]}{}^{a(k-1)c,b(m)} \nonumber \\
 && \qquad \qquad  + e_{[\mu}{}^{a}
\xi_{\nu]}{}^{a(k-1)b,b(m-1)c} - (k+1) e_{[\mu}{}^{b}
\xi_{\nu]}{}^{a(k),b(m-1)c} - Tr ] \label{gaugeb5}
\end{eqnarray}
They produce additional variations:
\begin{eqnarray*}
- 4(-1)^{k+m} a_{k,m} &\epthree& [ \frac{1}{k} 
\eta_\mu{}^{d(k),ae(m-1),b} \Psi_{\nu\alpha}{}^{d(k),ce(m-1)}
+ \frac{1}{m} \eta_\mu{}^{ad(k-1),e(m),b} 
\Psi_{\nu\alpha}{}^{cd(k-1),e(m)} \\ 
&& + \frac{1}{k} \Omega_{\mu\nu}{}^{d(k),ae(m-1)b}
\xi_\alpha{}^{d(k),ce(m-1)} + \frac{1}{m} 
\Omega_{\mu\nu}{}^{ad(k-1),e(m),b} \xi_\alpha{}^{cd(k-1),e(m)} ]
\end{eqnarray*}
As it can be easily seen it is not possible to restore the invariance
of the Lagrangian by adjusting the only parameter $a_{k,m}$. Similarly
to the previous case the invariance can be restored with the help of
additional fields only (see the next section).

\subsection{Special case --- tensor $Y(l+1,l+1)$}

This case corresponds to the rectangular Young tableau and also
deserves separate consideration. We need the two-forms 
$R_{\mu\nu}{}^{a(l),b(l)}$ and $\Omega_{\mu\nu}{}^{a(l),b(l),1}$ with
local indices corresponding to the Young tableau $Y(l,l)$ and 
$Y(l,l,1)$. The free Lagrangian in the flat case has the form:
\begin{eqnarray}
{\cal L}_0 &=& - \frac{1}{2} \epfour [ 
\Omega_{\mu\nu}{}^{ae(l-1),bf(l-1),g} 
\Omega_{\alpha\beta}{}^{ce(l-1),df(l-1),g} \nonumber \\
 && \qquad \qquad + \frac{2}{l} \Omega_{\mu\nu}{}^{e(l),af(l-1),b}
\Omega_{\alpha\beta}{}^{e(l),cf(l-1),d} ] \nonumber \\
 && + \epfive \Omega_{\mu\nu}{}^{af(l-1),bg(l-1),c} \partial_\alpha
R_{\beta\gamma}{}^{df(l-1),eg(l-1)} \label{lagb6}
\end{eqnarray}
It is invariant under the following gauge transformations:
\begin{eqnarray}
\delta_0 \Omega_{\mu\nu}{}^{a(l),b(l),c} &=& \partial_{[\mu}
\eta_{\nu]}{}^{a(l),b(l),c} \nonumber \\
\delta_0 R_{\mu\nu}{}^{a(l),b(l)} &=& \partial_{[\mu}
\xi_{\nu]}{}^{a(l),b(l)} + \eta_{[\mu}{}^{a(l),b(l),}{}_{\nu]}
\label{gaugeb6}
\end{eqnarray}

The replacement of the ordinary partial derivatives by the $AdS$
covariant ones produces
\begin{equation}
\delta {\cal L}_0 = \frac{4(l+2)(d+l-5)}{l} \kappa \epthree [
\Omega_{\mu\nu}{}^{d(l),ae(l-1),b} \xi_\alpha{}^{d(l),ce(l-1)} +
\eta_\mu{}^{d(l),ae(l-1),b} R_{\nu\alpha}{}^{d(l),ce(l-1)} ]
\end{equation}
But in this case the invariance of the Lagrangian can be restored
\cite{BMV00} with the help of introduction of mass-like terms:
\begin{equation}
{\cal L}_2 = a_{l,l} \epfour R_{\mu\nu}{}^{ae(l-1),bf(l-1)}
R_{\alpha\beta}{}^{ce(l-1),df(l-1)} \label{lagb7}
\end{equation}
as well as the corresponding corrections to the gauge transformations:
\begin{equation}
\delta_2 \Omega_{\mu\nu}{}^{a(l),b(l),c} = - 
\frac{2a_{l,l}}{(l+2)(d-4)} [ l e_{[\mu}{}^c 
\xi_{\nu]}{}^{a(l),b(l)} - e_{[\mu}{}^a \xi_{\nu]}{}^{a(l-1)c,b(l)}
- e_{[\mu}{}^b \xi_{\nu]}{}^{a(l),b(l-1)c} - Tr ] \label{gaugeb7}
\end{equation}
provided
$$
a_{l,l} = \frac{(l+2)(d+l-5)}{2}\kappa
$$

\section{Infinite spin bosonic fields}

In this section we consider the explicit Lagrangian descriptions for
the number of bosonic infinite spin cases. We begin with case based on
the completely symmetric tensors reproducing the results of Metsaev
\cite{Met16} but in the frame-like formalism. Then we present the
simplest example for the mixed symmetry tensors based on the so-called
long hooks. After that we turn to the case of the general mixed
symmetry tensors with two rows.

\subsection{Completely symmetric case --- $Y(k+1,0)$, $k \to \infty$}

We follow the results of Metsaev \cite{Met16} who has shown that
the same gauge invariant description for the massive higher spin
bosonic fields (\cite{Zin01} for the metric formulation and
\cite{Zin08b,PV10} for the frame-like one) is perfectly capable of
describing infinite spin massless and tahyonic cases provided one
abandons the restriction on the number of field components. Let us
recall that the main idea of such gauge invariant description is to
begin with the appropriate set of massless (finite component) fields
and then introduce the cross terms into Lagrangian as well as the
appropriate corrections to the gauge transformations gluing all them
together in such a way as to keep all the gauge symmetries intact and
as a result the number of physical degrees of freedom unchanged.

For the completely symmetric case we introduce a set of one-forms
($\omega_\mu{}^{a(k),b}$, $h_\mu{}^{a(k)}$), $1 \le k < \infty$, as
well as one-form $A_\mu$ and zero-forms $B^{ab}$, $\pi^a$ and
$\varphi$ for the spin 1 and spin 0 components. We begin with sum of
kinetic and mass-like terms for all fields:
\begin{eqnarray}
{\cal L} &=& \sum_{k=1}^\infty {\cal L}_0(h_\mu{}^{a(k)}) + 
\frac{1}{2} B_{ab}{}^2 - \eptwo \omega^{ab} D_\mu h_\nu - \frac{1}{2}
\pi_a{}^2 + \pi^\mu D_\mu \varphi \nonumber \\
 && + \sum_{k=1}^\infty {\cal L}_2(h_\mu{}^{a(k)}) + b_0 h \varphi +
c_0 \varphi^2
\end{eqnarray}
where ${\cal L}_0(h_\mu{}^{a(k)})$ and ${\cal L}_2(h_\mu{}^{a(k)})$
are defined in (\ref{lagb1}) and (\ref{lagb2}), and with the set of
initial gauge transformations defined in (\ref{gaugeb1}) and
(\ref{gaugeb2}) supplemented with
$$
\delta_0 h_\mu = D_\mu \zeta, \qquad \delta_2 \pi^a = b_0 \zeta^a
$$

Now we add a set of cross-terms gluing all these fields together (the
meaning of the coefficients $e_k$ is illustrated by Figure 1):
\begin{eqnarray}
{\cal L}_1 &=& \sum_{k=2}^\infty (-1)^k e_k \eptwo [ 
\omega_\mu{}^{ac(k-1),b} h_{\nu,c(k-1)} - h_\mu{}^{ac(k-1)}
\omega_{\nu,c(k-1)}{}^b ] \nonumber \\
 && - e_1 \eptwo \omega_\mu{}^{a,b} h_\nu + e_1 \omega^{\mu a}
h_{\mu,a} - e_0 \pi^\mu h_\mu
\end{eqnarray}
as well as the corresponding corrections to the gauge transformations:
\begin{eqnarray}
\delta_1 \omega_\mu{}^{a(k),b} &=& \frac{e_{k+1}}{2(k+1)} ((k+1) 
\chi_\mu{}^{a(k),b} + \chi^{a(k)b,}{}_\mu) + \frac{e_k}{2(d+k-2)} 
( e_\mu{}^a \chi^{a(k-1),b} - Tr ) \nonumber \\
\delta_1 h_\mu{}^{a(k)} &=& \frac{ke_{k+1}}{2(k+1)} \zeta_\mu{}^{a(k)}
+ \frac{e_k}{2(d+k-3)} (e_\mu{}^a \zeta^{a(k-1)} - \frac{2}{(d+2k-4)}
g^{a(2)} \zeta_\mu{}^{a(k-2)}) \\
\delta_1 \omega^{ab} &=& e_1 \chi^{ab}, \qquad
\delta_1 h_\mu = \frac{e_1}{2} \zeta_\mu, \qquad
\delta_1 \varphi = e_0 \zeta \nonumber
\end{eqnarray}

\setlength{\unitlength}{1mm}
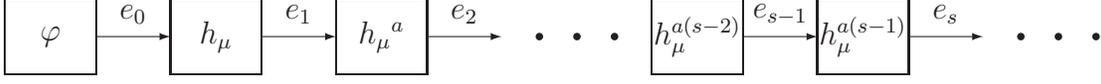
\begin{figure}
\begin{center}
\begin{picture}(150,12)

\put(1,1){\framebox(12,10)[]{$\varphi$}}
\put(13,6){\vector(1,0){10}}
\put(15,6){\makebox(6,6)[]{$e_0$}}

\put(23,1){\framebox(12,10)[]{$h_\mu$}}
\put(35,6){\vector(1,0){10}}
\put(37,6){\makebox(6,6)[]{$e_1$}}

\put(45,1){\framebox(12,10)[]{$h_\mu{}^a$}}
\put(57,6){\vector(1,0){10}}
\put(59,6){\makebox(6,6)[]{$e_2$}}

\multiput(72,6)(5,0){3}{\circle*{1}}

\put(87,1){\framebox(12,10)[]{$h_\mu^{a(s-2)}$}}
\put(99,6){\vector(1,0){10}}
\put(101,6){\makebox(6,6)[]{$e_{s-1}$}}

\put(109,1){\framebox(12,10)[]{$h_\mu^{a(s-1)}$}}
\put(121,6){\vector(1,0){10}}
\put(123,6){\makebox(6,6)[]{$e_s$}}

\multiput(136,6)(5,0){3}{\circle*{1}}

\end{picture}
\end{center}
\caption{Completely symmetric bosonic case}
\end{figure}

The requirement that the whole Lagrangian be invariant under all
gauge transformations produces a number of relations on the
parameters:
$$
(k+1)(d+k-1)b_{k+1} = k(d+k-2)b_k
$$
\begin{eqnarray*}
0 &=& \frac{(d+2k)(d+k-3)}{2(d+2k-2)(d+k-2)}e_{k+1}{}^2
- \frac{(k^2-1)}{2k^2}e_k{}^2 - \frac{2b_k}{k} 
+ \frac{2(k+1)(d+k-3)}{k} \kappa \\
0 &=& \frac{(d+2)(d-2)}{2d(d-1)}e_2{}^2 - e_1{}^2 - 2b_1 +
4(d-2)\kappa 
\end{eqnarray*}
$$
e_1{}^2 = \frac{2(d-2)}{d}c_0, \qquad
b_1 = - \frac{(d-2)}{2(d-1)}e_0{}^2, \qquad
b_0 = - \frac{e_1e_0}{2}
$$

The general solution of these equations has two free parameters and
their choice depends on  the case we are interested in. As we have
already mentioned, our system is capable of describing both massive
finite spin case (including all its massless and partially massless
limits) as well as the infinite spin (not necessarily massless) cases.
For the massive finite spin solution it is natural to choose the spin
and the mass as these two parameters where the spin $s$ is defined by
the requirement $e_s = 0$ (so that our chain of fields is restricted
from the above, see Figure 1), while we use the relation $m^2 =
(s-2)e_{s-1}{}^2/4$ as our definition of mass\footnote{As is well
known, there is no strict definition of mass and masslessness in
(Anti) de Sitter spaces. For the completely symmetric (spin-)tensors
it is natural to define the massless limit as the one where $e_{s-1}
\to 0$, i.e. the limit where the main gauge field completely decouples
from all the Stueckelberg ones. As for the concrete normalization of
mass, we choose it so that it coincides with the usual definition of
mass in flat Minkowski space.}. In these terms the solution looks
like:
\begin{eqnarray}
b_k &=& - \frac{s(d+s-3)}{k(d+k-2)} [ m^2 - (s-1)(d+s-4)\kappa]
\nonumber \\
e_k{}^2 &=& \frac{4(s-k)(d+s+k-3)}{(k-1)(d+2k-2)}
[ m^2 - (s-k-1)(d+s+k-4)\kappa ] \\
e_1{}^2 &=& \frac{4(s-1)(d+s-2)}{d} [ m^2 - (s-2)(d+s-3)\kappa ]
\nonumber \\
e_0{}^2 &=& \frac{2s(d+s-3)}{(d-2)} [ m^2 - (s-1)(d+s-4)\kappa]
\nonumber
\end{eqnarray}
Let us briefly recall the main properties of this solution. First of
all, from the expressions above it is clear that the massless limit $m
\to 0$ is possible in flat $\kappa = 0$ and Anti de Sitter $\kappa
< 0$ spaces only. In flat space massive spin-$s$ particle
decomposes into the whole set of massless fields with spins
$s,s-1,\dots,0$, while in $AdS$ space it decomposes into massless
spin-$s$ field and massive spin-$(s-1)$ one. In de Sitter space
$\kappa > 0$ we have a so-called unitary forbidden region $m^2 <
(s-1)(d+s-4)\kappa$. At the boundary of this region lives the first
partially massless case where the spin-0 component decouples. Inside
the forbidden  region we have a number of partially massless cases
where one of the $e_l = 0$. In this case the whole system decomposes
into two disconnected ones. The first one with the fields with $l \le
k
\le s-1$ describes unitary partially massless particle, while the
remaining fields gives non-unitary massive one.

Now let us turn to the main subject of our current work --- infinite
spin cases. Here we choose $c_0$ and $e_0$ as our two main parameters.
Then for all other parameters we obtain:
\begin{eqnarray}
b_k &=& - \frac{(d-2)}{2k(d+k-2)} e_0{}^2 \nonumber \\
e_k{}^2 &=& \frac{2}{(k-1)(d+2k-2)} [ k(d+k-3)c_0
- (k-1)(d+k-2) e_0{}^2 \nonumber \\
 && \qquad \qquad - 2k(k-1)(d+k-2)(d+k-3) \kappa ] \\
e_1{}^2 &=& \frac{2(d-2)}{d}c_0 \nonumber
\end{eqnarray}
As in the metric-like formulation by Metsaev \cite{Met16}, it is
convenient to introduce auxiliary variables $x_k = k(d+k-3)$. Then we
have
\begin{equation}
e_k{}^2 \sim - 2\kappa x_k{}^2 + [c_0-e_0{}^2+2(d-2)\kappa] x_k 
+ (d-2)e_0{}^2 \label{cond1}
\end{equation}
Now we consider three spaces in turn.

{\bf De Sitter space} From the expression above it is clear the 
$e_k{}^2$ will always become negative at some sufficiently large $k$
so that in general we obtain non-unitary theory. The only exceptions
appear each time when the parameters are adjusted so that some $e_s =
0$. These cases just reproduce massive finite spin particles described
above.

{\bf Minkowski space} Let us begin with  the $e_0{}^2 > 0$. If $c_0 <
e_0{}^2$ we again face the situation where $e_k{}^2$ become negative
at sufficiently large $k$ so that in general we obtain non-unitary
theory. And again, the only exceptions appear when we  adjust
parameters so that some $e_s = 0$ and these cases reproduce massive
finite spin particles. At $c_0 = e_0{}^2$ (that corresponds to the
$\mu_0 = 0$ in the Metsaev paper \cite{Met16}) we obtain unitary
massless infinite spin theory, while for the $c_0 > e_0{}^2$ we obtain
unitary tahyonic infinite spin one. A whole set of rather exotic
solutions arises if we assume that $e_0{}^2 < 0$ and again adjust the
parameters so that some $e_s = 0$:
$$
c_0 = \frac{(s-1)(d+s-2)}{s(d+s-3)} e_0{}^2
$$
These solutions resemble the massive finite spin case described above.
Indeed, the whole set of the fields decomposes into two disconnected
subsystems. But this time, the infinite chain with $s \le k < \infty$
describes unitary theory (which presumably corresponds to the
tahyonic infinite dimensional representation of the Poincar\'e
group), while the remaining fields correspond to the non-unitary
massive theory.

{\bf Anti de Sitter space} In this case (see Appendix A.2 and general
discussion in Appendix A.1) it is clear that $e_k{}^2$ will always
become positive at some sufficiently large $k$. So there are two
general possibilities. The first one is when all $e_k{}^2 > 0$ and we
obtain unitary infinite spin theory containing all fields with 
$0 \le k < \infty$. Besides, similarly to the flat case, there exist a
number of discrete solutions where some $e_s = 0$ and the unitary part
contains fields with $s \le k < \infty$ only.

\subsection{Mixed symmetry example --- $Y(k+1,1)$, $k \to \infty$}

In this section we present a simplest example for the mixed symmetry
tensors based on the so-called long hooks. We follow our previous
work \cite{Zin08c} for the massive finite component case but without
any restrictions on the number of field components. So we introduce
the following fields: $\Omega_\mu{}^{a(k),bc}$, 
$\Psi_{\mu\nu}{}^{a(k)}$, $\omega_\mu{}^{a(k),b}$ and 
$h_\mu{}^{a(k)}$, $1 \le k < \infty$ and $\Omega^{abc}$, 
$\Phi_{\mu\nu}$, $\omega^{ab}$ and $h_\mu$. The sum of the kinetic
terms (there are no possible mass-like terms in this particular
case):
\begin{eqnarray}
{\cal L} &=& \sum_{k=1}^\infty {\cal L}_0(\Phi_{\mu\nu}{}^{a(k)}) -
\Omega_{abc}{}^2 + \epthree \Omega^{abc} D_\mu \Phi_{\nu\alpha}
\nonumber \\
 && + \sum_{k=1}^\infty {\cal L}_0(h_\mu{}^{a(k)}) + \frac{1}{2}
\omega_{ab}{}^2 - \eptwo \omega^{ab} D_\mu h_\nu 
\end{eqnarray}
where Lagrangians ${\cal L}_0(\Phi_{\mu\nu}{}^{a(k)})$ and 
${\cal L}_0(h_\mu{}^{a(k)})$ are defined in (\ref{lagb3}) and
(\ref{lagb1}), with the initial set of the gauge transformations given
in (\ref{gaugeb3}) and (\ref{gaugeb1}) supplemented with
\begin{equation}
\delta_0 \Phi_{\mu\nu} = D_{[\mu} \xi_{\nu]}, \qquad
\delta_0 h_\mu = D_\mu \zeta
\end{equation}

There are three types of the possible cross-terms (see Figure 2). \\
$Y(k+1,1) \Leftrightarrow Y(k,1)$. In this case the cross-terms have
the form:
\begin{eqnarray}
{\cal L}_{11} &=& \sum_{k=2}^\infty (-1)^k c_k \epthree [ 
\Omega_\mu{}^{ae(k-1),bc} \Phi_{\nu\alpha,e(k-1)} +
\Omega_\mu{}^{e(k-1),ab} \Phi_{\nu\alpha,e(k-1)}{}^c ] \nonumber \\
 && - 3c_1 [ \eptwo \Omega^{abc} \Phi_{\nu\alpha,c} + \epthree
\Omega_\mu{}^{abc} \Phi_{\nu\alpha} ]
\end{eqnarray}
with the corresponding corrections to the gauge transformations:
\begin{eqnarray}
\delta_{11} \Omega_\mu{}^{a(k),bc} &=& - c_{k+1} [ 
\eta_\mu{}^{a(k),bc} - \frac{1}{(k+2)} \eta^{a(k)[b,c]}{}_\mu] 
 - \frac{c_k}{(d+k-2)} ( e_\mu{}^a \eta^{a(k-1),bc} - Tr) \nonumber \\
\delta_{11} \Phi_{\mu\nu}{}^{a(k)} &=& \frac{kc_{k+1}}{(k+2)}
\xi_{[\mu,\nu]}{}^{a(k)} - \frac{c_k}{(d+k-4)} (e_{[\mu}{}^a
\xi_{\nu]}{}^{a(k-1)} - Tr) \\
\delta_{11} \Phi_{\mu\nu}{}^a &=& - \frac{3c_1}{(d-3)} e_{[\mu}{}^a
\xi_{\nu]}, \qquad \delta_{11} \Omega^{abc} = - 3c_1 \eta^{abc},
\qquad \delta_{11} \Phi_{\mu\nu} = c_1 \xi_{[\mu,\nu]} \nonumber
\end{eqnarray}

\begin{figure}[htb]
\begin{center}
\begin{picture}(147,32)

\put(1,21){\framebox(12,10)[]{$\Phi_{\mu\nu}$}}
\put(6,21){\vector(0,-1){10}}
\put(7,13){\makebox(6,6)[]{$d_0$}}
\put(1,1){\framebox(12,10)[]{$h_\mu$}}

\put(13,26){\vector(1,0){10}}
\put(15,26){\makebox(6,6)[]{$c_1$}}
\put(13,6){\vector(1,0){10}}
\put(15,6){\makebox(6,6)[]{$e_1$}}

\put(23,21){\framebox(12,10)[]{$\Phi_{\mu\nu}^a$}}
\put(28,21){\vector(0,-1){10}}
\put(29,13){\makebox(6,6)[]{$d_1$}}
\put(23,1){\framebox(12,10)[]{$h_\mu^a$}}

\put(35,26){\vector(1,0){10}}
\put(37,26){\makebox(6,6)[]{$c_2$}}
\put(35,6){\vector(1,0){10}}
\put(37,6){\makebox(6,6)[]{$e_2$}}

\put(45,21){\framebox(12,10)[]{$\Phi_{\mu\nu}^{a(2)}$}}
\put(51,21){\vector(0,-1){10}}
\put(51,13){\makebox(6,6)[]{$d_2$}}
\put(45,1){\framebox(12,10)[]{$h_\mu^{a(2)}$}}

\put(57,26){\vector(1,0){10}}
\put(59,26){\makebox(6,6)[]{$c_3$}}
\put(57,6){\vector(1,0){10}}
\put(59,6){\makebox(6,6)[]{$e_3$}}

\multiput(71,16)(5,0){3}{\circle*{1}}

\put(87,21){\framebox(12,10)[]{$\Phi_{\mu\nu}^{a(s-2)}$}}
\put(92,21){\vector(0,-1){10}}
\put(94,13){\makebox(6,6)[]{$d_{s-2}$}}
\put(87,1){\framebox(12,10)[]{$h_\mu^{a(s-2)}$}}

\put(99,26){\vector(1,0){10}}
\put(101,26){\makebox(6,6)[]{$c_{s-1}$}}
\put(99,6){\vector(1,0){10}}
\put(101,6){\makebox(6,6)[]{$e_{s-1}$}}

\put(109,21){\framebox(12,10)[]{$\Phi_{\mu\nu}^{a(s-1)}$}}
\put(114,21){\vector(0,-1){10}}
\put(116,13){\makebox(6,6)[]{$d_{s-1}$}}
\put(109,1){\framebox(12,10)[]{$h_\mu^{a(s-1)}$}}

\put(121,26){\vector(1,0){10}}
\put(123,26){\makebox(6,6)[]{$c_s$}}
\put(121,6){\vector(1,0){10}}
\put(123,6){\makebox(6,6)[]{$e_s$}}

\multiput(136,16)(5,0){3}{\circle*{1}}

\end{picture}
\end{center}
\caption{Simplest mixed symmetry example}
\end{figure}
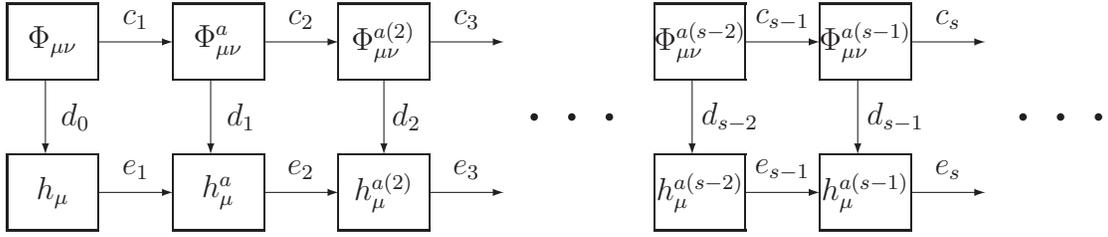

$Y(k+1,1) \Leftrightarrow Y(k+1,0)$. In this case we introduce
\begin{eqnarray}
{\cal L}_{12} &=& \sum_{k=1}^\infty (-1)^k d_k [ \frac{1}{k} \eptwo
\Omega_\mu{}^{e(k),ab} h_{\nu,e(k)} - \epthree 
\omega_\mu{}^{ae(k-1),b} \Phi_{\nu\alpha,e(k-1)}{}^c ] \nonumber \\
 && + d_0 \eptwo \omega^{ab} \Phi_{\mu\nu}
\end{eqnarray}
as well as the following corrections to the gauge transformations:
\begin{eqnarray}
\delta_{12} \Omega_\mu{}^{a(k),bc} &=& \frac{d_k}{2(k+2)(d-3)}
[ (k+1) \chi^{a(k),[b} e_\mu{}^{c]} + e_\mu{}^a \chi^{a(k-1)[b,c]} -
Tr ] \nonumber \\
\delta_{12} \Phi_{\mu\nu}{}^{a(k)} &=& - \frac{d_k}{2(k+2)(d+k-4)}
e_{[\mu}{}^a \zeta_{\nu]}{}^{a(k-1)} \\
\delta_{12} \omega_\mu{}^{a(k),b} &=& - \frac{d_k}{2} 
\eta^{a(k),b}{}_\mu, \qquad \delta_{12} h_\mu{}^{a(k)} = - d_k 
\xi_\mu{}^{a(k)}, \qquad \delta_{12} h_\mu = 2d_0 \xi_\mu \nonumber
\end{eqnarray}

$Y(k+1,0) \Leftrightarrow Y(k,0)$. This case has already been
considered in the previous subsection, so we do not repeat it here.

Now we require that the whole Lagrangian be invariant under all gauge
transformations. This gives us a number of relations on the
parameters:
$$
2(k+1)c_{k+1}d_k + (k+2)d_{k+1}e_{k+1} = 0
$$
$$
d_ke_{k+1} + \frac{2(d+k-2)}{(d+k-3)}c_{k+1}d_{k+1} = 0
$$
\begin{eqnarray*}
\frac{2k(d+k-4)(d+2k)}{(d+k-3)(d+2k-2)}c_{k+1}{}^2 &=&
\frac{2(k+2)(k-1)}{(k+1)} c_k{}^2 - d_k{}^2 - 2(k+2)(d+k-4)\kappa \\
\frac{2(d+2)(d-3)}{d(d-2)} c_2{}^2 &=& 18 c_1{}^2 - d_1{}^2 -
6(d-3)\kappa \\
\frac{k(d+k-3)(d+2k)}{2(d+k-2)(d+2k-2)} e_{k+1}{}^2 &=& 
\frac{(k+1)(k-1)}{2k} e_k{}^2 - \frac{(k+1)(d+k-3)}{(k+2)(d+k-4)}
d_k{}^2 \\
 && - 2(k+1)(d+k-3)\kappa
\end{eqnarray*}
$$
4c_1d_0 = d_1e_1, \qquad
3(d-2)c_1d_1 = (d-3)d_0e_1
$$
From the first two relations we obtain:
$$
(k+2)(d+k-2)d_{k+1}{}^2 = (k+1)(d+k-3)d_k{}^2
$$
$$
e_k{}^2 = \frac{4k(d+k-3)}{(k+1)(d+k-4)} c_k{}^2
$$

The general solution again depends on the two free parameters. For the
massive finite component $Y(s,1)$ case we choose $s$ and $m^2 =
\frac{sd_{s-1}{}^2}{2(s+1)}$ as our main ones\footnote{This time our
main gauge field $\Psi_{\mu\nu}{}^{a(s-1)}$ is connected with the two
Stueckelberg ones --- $\Psi_{\mu\nu}{}^{a(s-2)}$ and
$\Phi_\mu{}^{a(s-1)}$ with the coefficients $c_{s-1}$ and $d_{s-1}$
correspondingly (see Figure 2). These coefficients are related as
follows:
$$
2(s+1)(s-2)c_{s-1}{}^2 - sd_{s-1}{}^2 = 2s(s+1)(d+s-5)\kappa
$$
Thus for $\kappa \ne 0$ it is impossible to set both these two
parameters to 0 and obtain the massless limit. So our $m$ here is not
the mass, but simply a convenient parameter.}. Then we obtain:
\begin{eqnarray}
d_k{}^2 &=& \frac{2(s+1)(d+s-4)}{(k+1)(d+k-3)} m^2 \nonumber \\
c_k{}^2 &=& \frac{(s-k)(d+k-4)}{(k-1)(d+2k-2)}
[m^2 + (k+1)(d+k-4)\kappa] \\
c_1{}^2 &=& \frac{(s-1)(d+s-2)}{6d} [ m^2 + 2(d-3)\kappa ]
\nonumber
\end{eqnarray}
Let us briefly describe the main features of such theory. This time
the massless limit is possible in flat Minkowski space only. In
de Sitter space we may set $m = 0$ (and hence all $d_k =
0$). Then the whole system decomposes into two disconnected ones with
the fields $\Psi_{\mu\nu}$ and $\Phi_\mu$ respectively. The first one
provides an example of partially massless theory for the mixed tensor
$Y(s,1)$, while the second one is the partially massless theory for
the completely symmetric tensor. In turn, in Anti de Sitter space
we may set $c_{s-1} = 0$. In this limit  the main pair of fields 
($\Psi_{\mu\nu}{}^{a(s-1)}$, $\Phi_\mu{}^{a(s-1)}$) decouples
providing one more example of partially massless theory, while the
remaining fields correspond to the massive $Y(s-1,1)$ theory. Also
there exists a number of non-unitary partially massless limits which
appear each time when we set some $c_k = 0$ (and hence $e_k = 0$).

For the infinite spin case we choose $c_1$ and $d_0$ as our main
parameters (see Figure 2). Then for the other parameters we obtain:
\begin{eqnarray}
c_k{}^2 &=& \frac{1}{(k-1)(d+2k-2)} [ (k+1)(d+k-4) C
 - (k-1)(d+k-2) D \nonumber \\
 && \qquad \qquad - (k-1)(k+1)(d+k-2)(d+k-4)\kappa ] \\
d_k{}^2 &=& \frac{8(d-3)}{3(k+1)(d+k-3)} d_0{}^2 \nonumber 
\end{eqnarray}
where
\begin{equation}
C = \frac{3dc_1{}^2}{(d-3)}, \qquad
D = \frac{2d_0{}^2}{3}
\end{equation}
To analyze this solution it is convenient to introduce
$x_k = (k+1)(d+k-4)$. Then we have:
\begin{equation}
c_k{}^2 \sim - \kappa x_k{}^2 + [C - D + 2(d-3)\kappa] x_k + 2(d-3)D,
\qquad d_k{}^2 \sim D \label{cond2}
\end{equation}

{\bf De Sitter space} There are no unitary models except for the
massive finite component $Y(s,1)$ ones described above.

{\bf Flat Minkowski space} There are two general possibilities. For $C
= D > 0$ we obtain unitary massless infinite spin theory, while for $C
> D > 0$ we obtain unitary tahyonic infinite spin one. Note, that in
the second case there is a limit $D \to 0$ that resembles the
partially massless limit in de Sitter space. Indeed, all $d_k \to
0$ and the whole system decomposes into two disconnected ones with the
field $\Psi_{\mu\nu}$ and $\Phi_\mu$ respectively.

{\bf Anti de Sitter space} First of all there are solutions where all
$c_k{}^2 > 0$ so that we obtain unitary infinite spin theory
containing all our fields (see Appendix A.3). Moreover, it is possible
to take a limit $D \to 0$ (so that all $d_k \to 0$) when the whole
system decompose into two subsystems with the fields $\Psi_{\mu\nu}$
and $\Phi_\mu$ respectively. The second class of solutions appears
when some $c_s = 0$ (and hence $e_s = 0$), while all $c_k{}^2 > 0$ for
$k > s$. In this case the unitary part of the theory contains fields
$\Psi_{\mu\nu}{}^{a(k)}$ and $\Phi_\mu{}^{a(k)}$ with $k > s$. Note
that in this case it is also possible to take a limit $D \to 0$ so
that the whole system decomposes into four independent ones where the
subsystem  where the fields $\Psi_{\mu\nu}{}^{a(k)}$, 
$s \le k < \infty$ provides one more non-trivial example of the
unitary infinite spin theory.

\subsection{General case $Y(k+1,l+1)$, $k \to \infty$}

In this case we also follow our previous work on the massive
mixed symmetry tensors \cite{Zin09c} but without a restriction on the
number of field components. Thus we introduce:
($\Omega_{\mu\nu}{}^{a(k),b(m),c}$, 
$\Psi_{\mu\nu}{}^{a(k),b(m)}$), $l \le k < \infty$, $1 \le m \le l$,
($\Omega_\mu{}^{a(k),bc}$, $\Phi_{\mu\nu}{}^{a(k)}$) and
($\omega_\mu{}^{a(k),b}$, $h_\mu{}^{a(k)}$), $l \le k < \infty$.
The sum of kinetic and mass-like terms:
\begin{eqnarray}
{\cal L} &=& \sum_{k=l}^\infty [ \sum_{m=1}^l 
{\cal L}_0(\Psi_{\mu\nu}{}^{a(k),b(m)}) + 
{\cal L}_0(\Phi_{\mu\nu}{}^{a(k)}) + {\cal L}_0(h_\mu{}^{a(k)}) ]
\nonumber \\
 && + \sum_{k=l}^\infty [ \sum_{m=1}^l 
{\cal L}_2(\Psi_{\mu\nu}{}^{a(k),b(m)}) + b_{k,1} \epthree
\Psi_{\mu\nu}{}^{ae(k-1),b} h_\alpha{}^{ce(k-1)} +
{\cal L}_2(h_\mu{}^{a(k)}) ]
\end{eqnarray}
where the kinetic terms were defined in (\ref{lagb4}) ((\ref{lagb6})
for the special case $k=m=l$), (\ref{lagb3}) and (\ref{lagb1}), while
the mass-like terms in (\ref{lagb5}) ((\ref{lagb7}) for $k=m=l$) and 
(\ref{lagb2}). Also as the initial set of gauge transformations we
use expressions given at (\ref{gaugeb1}),
(\ref{gaugeb2}), (\ref{gaugeb3}), (\ref{gaugeb4}),
(\ref{gaugeb5}),(\ref{gaugeb6}), (\ref{gaugeb7}).

Now we consider possible cross-terms (see Figure 3).\\
$Y(k+1,m+1) \Leftrightarrow Y(k,m+1)$. In this case they have the
form:
\begin{eqnarray}
(-1)^{k+m} {\cal L}_1 &=& c_{k,m} \epfour [ 
\Omega_{\mu\nu}{}^{ae(k-1),bf(m-1),c} 
\Psi_{\alpha\beta}{}^{e(k-1),df(m-1)} \nonumber \\
 && \qquad \qquad \Omega_{\mu\nu}{}^{e(k-1),af(m-1),b}
\Psi_{\alpha\beta}{}^{ce(k-1),df(m-1)} ]
\end{eqnarray}
with the corresponding corrections for the gauge transformations
\begin{eqnarray}
\delta_1 \Omega_{\mu\nu}{}^{a(k),b(m),c} &=& - \frac{c_{k,m}}{(d+k-3)}
[ e_{[\mu}{}^a \eta_{\nu]}{}^{a(k-1),b(m),c} - Tr ] \nonumber \\
 && + \frac{kc_{k+1,m}}{(k+1)} [ \eta_{[\mu,\nu]}{}^{a(k),b(m),c}
+ \frac{1}{(k-m+2)} \eta_{[\mu}{}^{a(k)b,b(m-1)}{}_{\nu]} \nonumber \\
 && + \frac{1}{(k+2)} \eta_{[\mu}{}^{a(k)c,b(m)}{}_{\nu]} +
\frac{1}{(k+2)(k-m+2)} \eta_{[\mu}{}^{a(k)b,b(m-1)c}{}_{\nu]} ] \\
\delta_1 \Phi_{\mu\nu}{}^{a(k),b(m)} &=&  - \frac{c_{k,m}}{(d+k-4)}
[ e_{[\mu}{}^a \xi_{\nu]}{}^{a(k-1),b(m)} - Tr] \nonumber \\
 && + \frac{kc_{k+1,m}}{(k+2)} [ \xi_{[\mu,\nu]}{}^{a(k),b(m)} +
\frac{1}{(k-m+2)} \xi_{[\mu}{}^{a(k-1)b,b(m-1)}{}_{\nu]} ] \nonumber
\end{eqnarray}
Note that the two special cases $Y(k+1,1) \Leftrightarrow Y(k,1)$ and
$Y(k+1,0) \Leftrightarrow Y(k,0)$ have already been considered in the
previous subsections.

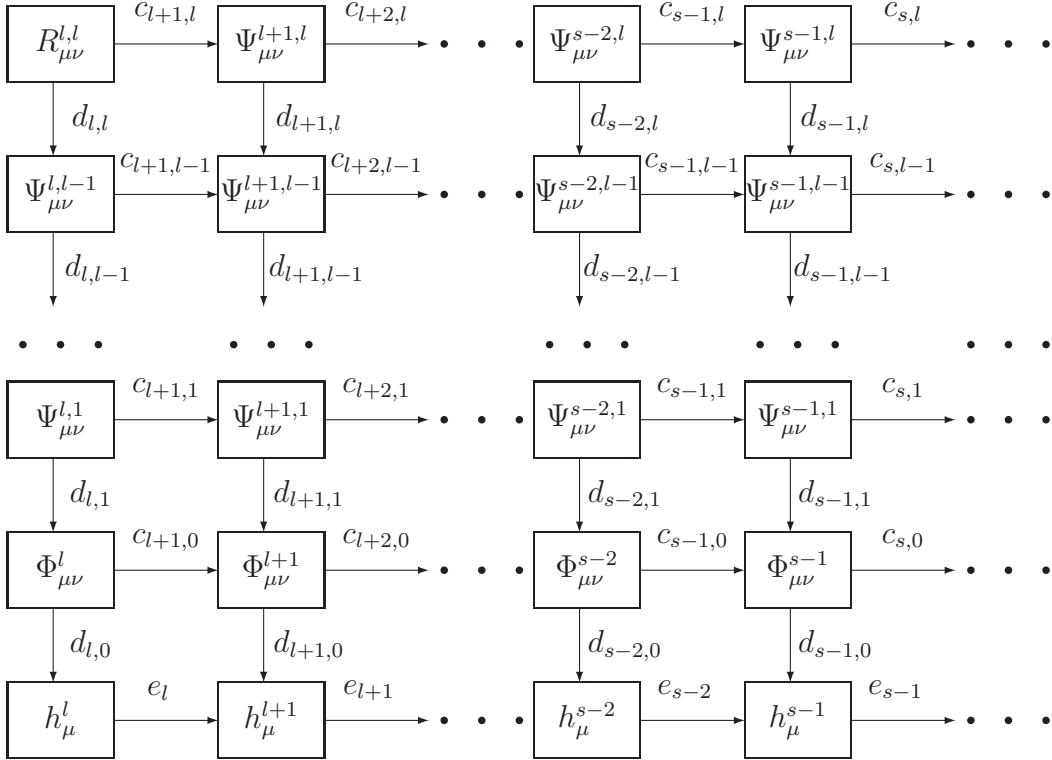
\begin{figure}[htb]
\begin{center}
\begin{picture}(140,102)

\put(1,91){\framebox(14,10)[]{$R_{\mu\nu}^{l,l}$}}
\put(15,96){\vector(1,0){14}}
\put(16,97){\makebox(12,6)[]{$c_{l+1,l}$}}
\put(7,91){\vector(0,-1){10}}
\put(8,83){\makebox(8,6)[]{$d_{l,l}$}}

\put(1,71){\framebox(14,10)[]{$\Psi_{\mu\nu}^{l,l-1}$}}
\put(15,76){\vector(1,0){14}}
\put(16,77){\makebox(12,6)[]{$c_{l+1,l-1}$}}
\put(7,71){\vector(0,-1){10}}
\put(8,63){\makebox(10,6)[]{$d_{l,l-1}$}}

\multiput(3,56)(5,0){3}{\circle*{1}}

\put(1,41){\framebox(14,10)[]{$\Psi_{\mu\nu}^{l,1}$}}
\put(15,46){\vector(1,0){14}}
\put(16,47){\makebox(12,6)[]{$c_{l+1,1}$}}
\put(7,41){\vector(0,-1){10}}
\put(8,33){\makebox(8,6)[]{$d_{l,1}$}}

\put(1,21){\framebox(14,10)[]{$\Phi_{\mu\nu}^l$}}
\put(15,26){\vector(1,0){14}}
\put(16,27){\makebox(12,6)[]{$c_{l+1,0}$}}
\put(7,21){\vector(0,-1){10}}
\put(8,13){\makebox(8,6)[]{$d_{l,0}$}}

\put(1,1){\framebox(14,10)[]{$h_\mu^l$}}
\put(15,6){\vector(1,0){14}}
\put(16,7){\makebox(10,6)[]{$e_l$}}

\put(29,91){\framebox(14,10)[]{$\Psi_{\mu\nu}^{l+1,l}$}}
\put(43,96){\vector(1,0){14}}
\put(44,97){\makebox(12,6)[]{$c_{l+2,l}$}}
\put(35,91){\vector(0,-1){10}}
\put(36,83){\makebox(10,6)[]{$d_{l+1,l}$}}

\put(29,71){\framebox(14,10)[]{$\Psi_{\mu\nu}^{l+1,l-1}$}}
\put(43,76){\vector(1,0){14}}
\put(44,77){\makebox(12,6)[]{$c_{l+2,l-1}$}}
\put(35,71){\vector(0,-1){10}}
\put(36,63){\makebox(12,6)[]{$d_{l+1,l-1}$}}

\multiput(31,56)(5,0){3}{\circle*{1}}

\put(29,41){\framebox(14,10)[]{$\Psi_{\mu\nu}^{l+1,1}$}}
\put(43,46){\vector(1,0){14}}
\put(44,47){\makebox(12,6)[]{$c_{l+2,1}$}}
\put(35,41){\vector(0,-1){10}}
\put(36,33){\makebox(10,6)[]{$d_{l+1,1}$}}

\put(29,21){\framebox(14,10)[]{$\Phi_{\mu\nu}^{l+1}$}}
\put(43,26){\vector(1,0){14}}
\put(44,27){\makebox(12,6)[]{$c_{l+2,0}$}}
\put(35,21){\vector(0,-1){10}}
\put(36,13){\makebox(10,6)[]{$d_{l+1,0}$}}

\put(29,1){\framebox(14,10)[]{$h_\mu^{l+1}$}}
\put(43,6){\vector(1,0){14}}
\put(44,7){\makebox(10,6)[]{$e_{l+1}$}}

\multiput(59,96)(5,0){3}{\circle*{1}}
\multiput(59,76)(5,0){3}{\circle*{1}}
%\multiput(59,56)(5,0){3}{\circle*{1}}
\multiput(59,46)(5,0){3}{\circle*{1}}
\multiput(59,26)(5,0){3}{\circle*{1}}
\multiput(59,6)(5,0){3}{\circle*{1}}

\put(71,91){\framebox(14,10)[]{$\Psi_{\mu\nu}^{s-2,l}$}}
\put(85,96){\vector(1,0){14}}
\put(86,97){\makebox(12,6)[]{$c_{s-1,l}$}}
\put(77,91){\vector(0,-1){10}}
\put(78,83){\makebox(10,6)[]{$d_{s-2,l}$}}

\put(71,71){\framebox(14,10)[]{$\Psi_{\mu\nu}^{s-2,l-1}$}}
\put(85,76){\vector(1,0){14}}
\put(86,77){\makebox(12,6)[]{$c_{s-1,l-1}$}}
\put(77,71){\vector(0,-1){10}}
\put(78,63){\makebox(12,6)[]{$d_{s-2,l-1}$}}

\multiput(73,56)(5,0){3}{\circle*{1}}

\put(71,41){\framebox(14,10)[]{$\Psi_{\mu\nu}^{s-2,1}$}}
\put(85,46){\vector(1,0){14}}
\put(86,47){\makebox(12,6)[]{$c_{s-1,1}$}}
\put(77,41){\vector(0,-1){10}}
\put(78,33){\makebox(10,6)[]{$d_{s-2,1}$}}

\put(71,21){\framebox(14,10)[]{$\Phi_{\mu\nu}^{s-2}$}}
\put(85,26){\vector(1,0){14}}
\put(86,27){\makebox(12,6)[]{$c_{s-1,0}$}}
\put(77,21){\vector(0,-1){10}}
\put(78,13){\makebox(10,6)[]{$d_{s-2,0}$}}

\put(71,1){\framebox(14,10)[]{$h_\mu^{s-2}$}}
\put(85,6){\vector(1,0){14}}
\put(86,7){\makebox(10,6)[]{$e_{s-2}$}}

\put(99,91){\framebox(14,10)[]{$\Psi_{\mu\nu}^{s-1,l}$}}
\put(113,96){\vector(1,0){14}}
\put(114,97){\makebox(12,6)[]{$c_{s,l}$}}
\put(105,91){\vector(0,-1){10}}
\put(106,83){\makebox(10,6)[]{$d_{s-1,l}$}}

\put(99,71){\framebox(14,10)[]{$\Psi_{\mu\nu}^{s-1,l-1}$}}
\put(113,76){\vector(1,0){14}}
\put(114,77){\makebox(12,6)[]{$c_{s,l-1}$}}
\put(105,71){\vector(0,-1){10}}
\put(106,63){\makebox(12,6)[]{$d_{s-1,l-1}$}}

\multiput(101,56)(5,0){3}{\circle*{1}}

\put(99,41){\framebox(14,10)[]{$\Psi_{\mu\nu}^{s-1,1}$}}
\put(113,46){\vector(1,0){14}}
\put(114,47){\makebox(12,6)[]{$c_{s,1}$}}
\put(105,41){\vector(0,-1){10}}
\put(106,33){\makebox(10,6)[]{$d_{s-1,1}$}}

\put(99,21){\framebox(14,10)[]{$\Phi_{\mu\nu}^{s-1}$}}
\put(113,26){\vector(1,0){14}}
\put(114,27){\makebox(12,6)[]{$c_{s,0}$}}
\put(105,21){\vector(0,-1){10}}
\put(106,13){\makebox(10,6)[]{$d_{s-1,0}$}}

\put(99,1){\framebox(14,10)[]{$h_\mu^{s-1}$}}
\put(113,6){\vector(1,0){14}}
\put(114,7){\makebox(10,6)[]{$e_{s-1}$}}

\multiput(129,96)(5,0){3}{\circle*{1}}
\multiput(129,76)(5,0){3}{\circle*{1}}
\multiput(129,56)(5,0){3}{\circle*{1}}
\multiput(129,46)(5,0){3}{\circle*{1}}
\multiput(129,26)(5,0){3}{\circle*{1}}
\multiput(129,6)(5,0){3}{\circle*{1}}

\end{picture}
\end{center}
\caption{General mixed symmetry case}
\end{figure}

$Y(k+1,m+1) \Leftrightarrow Y(k+1,m)$. here we introduce:
\begin{eqnarray}
(-1)^{k+m} {\cal L}_1 &=& d_{k,m} \epfour [ 
\Omega_{\mu\nu}{}^{ae(k-1),bf(m-1),c} 
\Psi_{\alpha\beta}{}^{de(k-1),f(m-1)} \nonumber \\
 && \qquad \qquad + \Omega_{\mu\nu}{}^{ae(k-1),fb(m-1),b}
\Psi_{\alpha\beta}{}^{ce(k-1),df(m-1)} ]
\end{eqnarray}
together with the following corrections:
\begin{eqnarray}
\delta_1 \Omega_{\mu\nu}{}^{a(k),b(m),c} &=&  
\frac{d_{k,m}}{(k-m+2)(d+m-4)} [ (k-m+1) e_{[\mu}{}^b
\eta_{\nu]}{}^{a(k),b(m-1),c} \nonumber \\
 && \qquad \qquad - e_{[\mu}{}^a \eta_{\nu]}{}^{a(k-1)b,b(m-1)c} -
Tr ] \nonumber \\
 && - d_{k,m+1} [ \eta_{[\mu}{}^{a(k),b(m)}{}_{\nu}{}^c + 
\frac{1}{(m+1)} \eta_{[\mu}{}^{a(k),b(m)c}{}_{\nu]} ] \nonumber \\
\delta_1 \Psi_{\mu\nu}{}^{a(k),b(m)} &=& - 
\frac{d_{k,m}}{(k-m+2)(d+m-5)} [ (k-m+1) \xi_{[\mu}{}^{a(k),b(m-1)}
e_{\nu]}{}^b \\
 && \qquad \qquad - e_{[\nu}{}^a \xi_{\mu]}{}^{a(k-1)b,b(m-1)} - Tr
] \nonumber \\
 && - \frac{m}{(m+1)} d_{k,m+1} \xi_{[\mu}{}^{a(k),b(m)}{}_{\nu]}
\nonumber
\end{eqnarray}
Here there is also a special case $Y(k+1,1) \Leftrightarrow Y(k+1,0)$
that has already been considered in the previous subsection.

The requirement that the whole Lagrangian be invariant under the all
appropriately corrected gauge transformations produces a great number
of relations on the parameters. First of all we obtain the following
important relations:
\begin{eqnarray*}
a_{k,m} &=& \frac{l(l+1)(d+l-3)(d+l-4)}{m(k+1)(d+k-3)(d+m-4)} a_{l,l}
\\
d_{k,m}{}^2 &=& \frac{(l-m+1)(d+l+m-3)}{(k-m+1)(d+k+m-3)} d_{l,m}{}^2
\\
c_{k,m}{}^2 &=& \frac{(k-l)(d+k+l-3)}{(k-m)(d+k+m-3)} c_{k,l}{}^2
\end{eqnarray*}
So all the parameters in the Lagrangian and gauge transformations are
expressed in terms of the main ones: $a_{l,l}$, $d_{l,m}$,
$m \le l$ (corresponding to the left-most column in Figure 3) and
$c_{k,l}$, $k > l$ (corresponding to the upper row). 

For these main parameters we obtain:
\begin{eqnarray*}
0 &=& \frac{2k(d+2k)((d+k+m-3)(d+k-4)-1)}{k(d+k-3)(d+2k-2)(d+k+m-3)} 
c_{k+1,m}{}^2 \\
 && - \frac{2(k-1)((k+2)(k-m+1)-1)}{k(k+1)(k-m+1)} c_{k,m}{}^2 \\
 && - \frac{2m(d+2m-2)((d+k+m-3)(d+m-5)+(k-m+1)(m+2)-1)}
{(k-m+1)(d+2m-4)(d+m-4)(d+k+m-3)} d_{k,m+1}{}^2 \\
 && - \frac{2(m-1)}{km} d_{k,m}{}^2
- \frac{4}{k} a_{k,m}
+ \frac{2(k+2)(d+k-4)+(d+m-6)}{k} \kappa
\end{eqnarray*}
\begin{eqnarray*}
0 &=& - \frac{2k(d+2k)(m+1)}{m(d+k-3)(d+2k-2)(d+k+m-3)} 
c_{k+1,m}{}^2 \\
 && + \frac{2(k-1)(m+1)}{m(k+1)(k-m+1)} c_{k,m}{}^2 \\
 && + \frac{2(d+2m-2)((k-m+1)(d+k+m-3)(d+m-5)+m+1)}
{(k+m-1)(d+2m-4)(d+m-4)(d+k+m-3)} d_{k,m+1}{}^2 \\
 && - \frac{2(m-1)(m+1)}{m^2} d_{k,m}{}^2
- 4 \frac{1}{m} a_{k,m}
+ 2 \frac{(m+1)(d+m-6)}{m} \kappa
\end{eqnarray*}

Once again the general solution of these equations depends on the two
arbitrary parameters and it is capable of describing both the massive
finite component case as well as a number of infinite spin cases.

For the massive case defined with $c_{s,l} = 0$ (and hence all
$c_{s,m} = 0$, $0 \le m \le l$) we obtain\footnote{As in the previous
subsection our main gauge field $\Psi_{\mu\nu}{}^{a(s-1),b(l)}$ is
connected with the two Stueckelberg ones 
$\Psi_{\mu\nu}{}^{a(s-2),b(l)}$ and $\Psi_{\mu\nu}{}^{a(s-1),b(l-1)}$
with the parameters $c_{s-1,l}$ and $d_{s-1,l}$ respectively (see
Figure 3). The relation on these parameters
$$
c_{s-1,l}{}^2 - d_{s-1,l}{}^2 \sim \kappa
$$
shows that for $\kappa \ne 0$ it is impossible to set both these
parameters to 0 and obtain the massless limit. Thus our $M$ here is
not the mass, but just a convenient parameter.}:
\begin{eqnarray*}
a_{l,l} &=& - \frac{(s+1)(d+s-4)}{2l(d+l-3)} [ M^2 - l(d+l-5)\kappa ]
\\
c_{k,l}{}^2 &=& \frac{(s-k)(d+s+k+1)}{(k-1)(d+2k-2)}
[ M^2 + (k-l+1)(d+k+l-4)\kappa] \\
d_{l,m}{}^2 &=& \frac{(s-m+1)(d+l+m-4)(d+s+m-4)}
{(m-1)(d+2m-4)(d+l+m-3)} [ M^2 - (l-m)(d+l+m-5)\kappa] \\
d_{l,1}{}^2 &=& \frac{s(d+l-3)(d+s-3)}{(d-2)(d+l-2)}
[ M^2 - (l-1)(d+l-4)\kappa] \\
d_{l,0}{}^2 &=& \frac{2(s+1)(d+l-4)(d+s-4)}{(d-4)(d+l-3)}
[ M^2 - l(d+l-5)\kappa ]
\end{eqnarray*}
In de Sitter space we have an unitary forbidden region $M^2 <
l(d+l-5)\kappa$. At the boundary of this region we find the only
unitary partially massless limit where $d_{l,0} = 0$ (and hence all
$d_{k,0} = 0$) and all the fields in the lowest row on Figure 3
decouple from the rest ones. Inside the forbidden region there is a
number of non-unitary partially massless limits which appear each time
when some $d_{l,m} = 0$ (and hence all $d_{k,m} = 0$). In this case
the Figure 3 splits vertically into two disconnected parts.

In Anti de Sitter space we also have an unitary forbidden region
$M^2 < - (s-l)(d+s+l-5)\kappa$. At the boundary of this region we
again find the only unitary partially massless limit where $c_{s-1,l}
= 0$ (and hence all $c_{s-1,m} = 0$) so that the right-most column in
Figure 3 decouples. Inside the forbidden region there exists a number
of non-unitary partially massless limit which appear each time when
some $c_{k,l} = 0$ (and hence all $c_{k,m} = 0$) when the Figure 3
splits horizontally.

For the infinite spin case we choose $c_{l+1,l}$ and $d_{l,l}$ as our
main parameters (see Figure 3). Then we obtain:
\begin{eqnarray*}
a_{l,l} &=& \frac{(d+l-5)}{4(d+l-3)} C - \frac{(l+2)}{4l} D
+ \frac{(l+2)(d+l-5)}{2} \kappa \\
c_{k,l}{}^2 &=& \frac{1}{2(k-1)(d+2k-2)} 
 [ (k-l+1)(d+k+l-4) C \\
 && - (k-l-1)(d+k+l-2) D \\
 && - 2(k-l+1)(k-l-1)(d+k+l-2)(d+k+l-4)\kappa ] \\
d_{l,k}{}^2 &=& \frac{(d+l+k-4)}{2(k-1)(d+2k-4)(d+l+k-3)}
[ - (l-k)(d+l+k-5) C \\
 && + (l-k+2)(d+l+k-3) D \\
 && - 2(l-k)(l-k+2)(d+l+k-3)(d+l+k-5)\kappa ]
\end{eqnarray*}
where
$$
C = \frac{l(d+2l)}{(d+2l-3)}c_{l+1,l}{}^2,
\qquad D = (l-1)d_{l,l}{}^2
$$
Let us introduce the variables $x_k = k(d+k-5)$ and the function
\begin{eqnarray}
F(x_k) &=& - 2\kappa x_k{}^2 + [ C - D + (l(d+l-5) + (l+2)(d+l-3))
\kappa] x_k \nonumber \\
 && + (l+2)(d+l-3)D - l(d+l-5)C - 2l(l+2)(d+l-3)(d+l-5)\kappa
\label{cond3}
\end{eqnarray}
Then we find:
\begin{eqnarray*}
c_{k-1,l}{}^2 &\sim& F(x_k), \qquad l+2 \le k < \infty \\
d_{l,k}{}^2 &\sim& F(x_k), \qquad 0 \le k \le l \\
a_{l,l} &\sim& F(x_0)
\end{eqnarray*}
In flat Minkowski space we obtain a region of the parameters
$$
C \ge D > \frac{l(d+l-5)}{(l+2)(d+l-3)} C
$$
corresponding to the unitary infinite spin theory containing all our
fields. Beyond this region there exists a number of partially massless
limits where one of the $d_{l,k} = 0$ (and hence all $d_{m,k} = 0$, $m
\ge l$):
$$
D = \frac{(l-k)(d+l+k-5)}{(l-k+2)(d+l+k-3)}C, \qquad C > 0
$$
and the Figure 3 splits vertically into two disconnected parts.

In Anti de Sitter space there exists a whole region of parameters
(see Appendix A.4) which corresponds to the unitary infinite spin
theory containing all our fields. Beyond this region there are two
types of discrete solutions. The first one corresponds to the cases
where some $c_{s,l} = 0$ (and hence all $c_{s,m} = 0$, $0 \le m \le
l$) so that Figure 3 splits horizontally. The second one corresponds
to the cases where some $d_{l,m} = 0$ (and hence all $d_{k,m} = 0$, $l
\le k < \infty$) when Figure 3 splits vertically. Moreover, these two
sets have common points when simultaneously some $c_{s,l} = 0$ and
$d_{l,l} = 0$ so that Figure 3 splits into the four disconnected
parts.

\section{Massless fermionic fields}

In this section we provide all necessary information on the massless
(finite component) symmetric and mixed symmetry fermionic fields which
will serve as the building blocks for our construction of the
infinite component cases. In what follows $Y(k+1/2,l+1/2)$ 
denote mixed symmetry spin-tensor having two rows of tensor indices
with length $k$ and $l$ respectively as well as implicit spinor index.

\subsection{Completely symmetric spin-tensor $Y(k+3/2,1/2)$}

The frame-like description for such spin-tensor \cite{Vas88} uses the
one-form $\Phi_\mu{}^{a(k)}$, completely symmetric on its local
indices and $\gamma$-transverse, i.e. $\gamma_b \Phi_\mu{}^{ba(k-1)} =
0$. The free Lagrangian in the flat case has the form:
\begin{equation}
(-1)^k {\cal L}_0(\Phi_\mu{}^{a(k)}) = -i \epthree [ 
\bar{\Phi}_\mu{}^{d(k)} \Gamma^{abc} \partial_\nu \Phi_\alpha{}^{d(k)}
- 6k \bar{\Phi}_\mu{}^{ad(k-1)} \gamma^b \partial_\nu 
\Phi_\alpha{}^{cd(k-1)} ] \label{lagf1}
\end{equation}
It is invariant under the following gauge transformations:
\begin{equation}
\delta_0 \Phi_\mu{}^{a(k)} = \partial_\mu \zeta^{a(k)} + 
\eta^{a(k),}{}_\mu \label{gaugef1}
\end{equation}
where parameters $\zeta^{a(k)}$ and $\eta^{a(k),b}$ are such that:
$$
(\gamma\zeta)^{a(k-1)} = 0, \qquad
\eta^{a(k),a} = 0, \qquad 
(\gamma\eta)^{a(k-1),b} = \gamma_b \eta^{a(k),b} = 0
$$
If we replace the ordinary partial derivatives by the $AdS$ covariant
ones, the Lagrangian ceases to be gauge invariant:
$$
\delta {\cal L}_0 = i(-1)^k \frac{3(d+2k-1)(d+2k-2)}{2} \kappa
e^\mu{}_a \bar{\Phi}_\mu{}^{b(k)} \gamma^a \zeta_{b(k)}
$$
But the gauge invariance of the Lagrangian can be restored by adding
the following mass-like terms to the Lagrangian:
\begin{equation}
{\cal L}_1 = (-1)^{k+1} b_k \eptwo [ \bar{\Phi}_\mu{}^{c(k)}
\Gamma^{ab} \Phi_\nu{}^{c(k)} + 2k \bar{\Phi}_\mu{}^{ac(k-1)} 
\Phi_\nu{}^{bc(k-1)} ] \label{lagf2}
\end{equation}
as well as the corresponding corrections to the gauge transformations
\begin{equation}
\delta_1 \Phi_\mu{}^{a(k)} = - \frac{ib_k}{3(d-2)} [ \gamma_\mu
\zeta^{a(k)} - \frac{2}{(d+2k-2)} \gamma^a \zeta_\mu{}^{a(k-1)} ]
\label{gaugef2}
\end{equation}
provided
$$
b_k{}^2 = - \frac{9}{4}(d+2k-2)^2 \kappa
$$
As usual for the fermions it requires $\kappa < 0$, i.e $AdS$ space.

\subsection{Mixed symmetry spin-tensor $Y(k+3/2,3/2)$}

In this and the following subsections we use the results of
\cite{Zin09b} (see also \cite{Zin09a,SZ10}) which provided the
generalization of the bosonic formalism \cite{Skv08} to the fermionic
case.

Let us consider the simplest example of the mixed symmetric 
spin-tensors which appears to be special and has to be considered
separately. We need the two-form $\Psi_{\mu\nu}{}^{a(k)}$,
completely symmetric and $\gamma$-transverse on its local indices. The
free Lagrangian in the flat space can be written as follows:
\begin{equation}
(-1)^k {\cal L}_0(\Psi_{\mu\nu}{}^{a(k)}) = - i \epfive [ 
\bar{\Psi}_{\mu\nu}{}^{f(k)} \Gamma^{abcde} \partial_\alpha 
\Psi_{\beta\gamma}{}^{f(k)} - 10k \bar{\Psi}_{\mu\nu}{}^{af(k-1)}
\Gamma^{bcd} \partial_\alpha \Psi_{\beta\gamma}{}^{ef(k-1)} ]
\label{lagf3}
\end{equation}
It is invariant under the following gauge transformations:
\begin{equation}
\delta_0 \Psi_{\mu\nu}{}^{a(k)} = \partial_{[\mu} \xi_{\nu]}{}^{a(k)}
+ \eta^{a(k),}{}_{\mu\nu} \label{gaugef3}
\end{equation}
As usual, the replacement of the ordinary partial derivatives spoils
the invariance of the Lagrangian that we can try to restore by
introducing the mass-like terms to the Lagrangian
\begin{equation}
{\cal L}_1 = (-1)^k a_k \epfour [ \bar{\Psi}_{\mu\nu}{}^{e(k)}
\Gamma^{abcd} \Psi_{\alpha\beta}{}^{e(k)} + 6k 
\bar{\Psi}_{\mu\nu}{}^{ae(k-1)} \Gamma^{bc} 
\Psi_{\alpha\beta}{}^{de(k-1)} ] \label{lagf4}
\end{equation}
as well as the corresponding corrections to the gauge transformations:
\begin{equation}
\delta_1 \Psi_{\mu\nu}{}^{a(k)} = \frac{ia_k}{5(d-4)} [
\gamma_{\mu} \xi_{\nu]}{}^{a(k)} + \frac{2}{(d+2k-2)} \gamma^a
\xi_{[\mu,\nu]}{}^{a(k-1)} ] \label{gaugef4}
\end{equation}
Collecting together all variations we obtain
$$
i (-1)^k [ - \frac{8(d-3)}{5(d-4)} a_{k,0}{}^2 - 10(d-3)(d-4) \kappa ]
\epthree [ \bar{\Psi}_{\mu\nu}{}^{(k)} \Gamma^{abc} \xi_\alpha{}^{(k)}
- 3 k \bar{\Psi}_{\mu\nu}{}^{a(k-1)} \gamma^b \xi_\alpha{}^{c(k-1)} ]
$$
$$
i (-1)^k [ \frac{8k a_{k,0}{}^2}{5(d-4)(d+2k-2)} - 10k(2d+2k-5) \kappa
] \epthree [ \bar{\Psi}_{\mu\nu}{}^{(k)} \Gamma^{abc} 
\xi_\alpha{}^{(k)} + 6 \bar{\Psi}_{\mu\nu}{}^{a(k-1)} \gamma^b 
\xi_\alpha{}^{c(k-1)} ]
$$
As it is easy to see it is impossible to restore the gauge invariance
by adjusting the only parameter $a_{k,0}$.

\subsection{General case --- spin-tensor $Y(k+3/2,m+3/2)$}

The frame-like formalism requires the two-form 
$\Psi_{\mu\nu}{}^{a(k),b(m)}$, with the local indices corresponding to
the Young tableau $Y(k,m)$. The free Lagrangian in the flat space
looks as follows:
\begin{eqnarray}
(-1)^{k+m} {\cal L}_0(\Psi_{\mu\nu}{}^{a(k),b(m)}) &=& - i \epfive
[ \bar{\Psi}_{\mu\nu}{}^{f(k),h(m)} \Gamma^{abcde} \partial_\alpha
\Psi_{\beta\gamma}{}^{f(k),h(m)} \nonumber \\
 && \qquad - 10k \bar{\Psi}_{\mu\nu}{}^{af(k-1),h(m)} \Gamma^{bcd} 
\partial_\alpha \Psi_{\beta\gamma}{}^{ef(k-1),h(m)} \nonumber \\
 && \qquad - 10m \bar{\Psi}_{\mu\nu}{}^{f(k),ah(m-1)} \Gamma^{bcd} 
\partial_\alpha \Psi_{|beta\gamma}{}^{f(k),eh(m-1)} \nonumber \\
 && \qquad - 60km  \bar{\Psi}_{\mu\nu}{}^{af(k-1),bh(m-1)} \gamma^c 
\partial_\alpha \Psi_{\beta\gamma}{}^{df(k-1),eh(m-1)} ] \label{lagf5}
\end{eqnarray}
It is invariant under the following gauge transformations:
\begin{equation}
\delta_0 \Psi_{\mu\nu}{}^{a(k),b(m)} = \partial_{[\mu}
\xi_{\nu]}{}^{a(k),b(m)} + \eta_{[\mu}{}^{a(k),b(m),}{}_{\nu]}
\label{gaugef5}
\end{equation}
The replacement of the ordinary partial derivatives by the $AdS$
covariant ones produces:
\begin{eqnarray}
\delta {\cal L}_0 &=& - 10 i \kappa (-1)^{k+m} \epthree \nonumber \\
&& [ [ (d+2k-3)(d+2m-4) + (k-m)(2k-2m+3)]
\bar{\Psi}_{\mu\nu}{}^{d(k),e(m)} \Gamma^{abc} 
\xi_\alpha{}^{d(k),e(m)} \nonumber \\
&& - 3 k [ (d+2m-5)(d+2m-6) + 3m - 4k -6]
\bar{\Psi}_{\mu\nu}{}^{ad(k-1),e(m)} \gamma^b 
\xi_\alpha{}^{cd(k-1),e(m)} \nonumber \\
&& - 3 m (d+2k-1)(d+2k-2)
\bar{\Psi}_{\mu\nu}{}^{d(k),ae(m-1)} \gamma^b
\xi_\alpha{}^{d(k),ce(m-1)} ]
\end{eqnarray}
At the same time if we introduce the mass-like term into the
Lagrangian:
\begin{eqnarray}
(-1)^{k+m} {\cal L}_1 &=& a_{k,m} \epfour [ 
\bar{\Psi}_{\mu\nu}{}^{f(k),h(m)} \Gamma^{abcd} 
\Psi_{\alpha\beta}{}^{f(k),h(m)} \nonumber \\
 && \qquad + 6k \bar{\Psi}_{\mu\nu}{}^{af(k-1),h(m)} \Gamma^{bc}
\Psi_{\alpha\beta}{}^{df(k-1),h(m)} \nonumber \\
 && \qquad + 6m \bar{\Psi}_{\mu\nu}{}^{f(k),ah(m-1)} \Gamma^{bc}
\Psi_{\alpha\beta}{}^{f(k),dh(m-1)} \nonumber \\
 && \qquad - 12km \bar{\Psi}_{\mu\nu}{}^{af(k-1),bh(m-1)}
\Psi_{\alpha\beta}{}^{cf(k-1),dh(m-1)} ] \label{lagf6}
\end{eqnarray}
as well as the corresponding corrections to the gauge transformations:
\begin{eqnarray}
\delta_1 \Psi_{\mu\nu}{}^{a(k),b(m)} &=& \frac{ia_{k,m}}{5(d-4)}
[ \gamma_{[\mu} \xi_{\nu]}{}^{a(k),b(m)} + \frac{2}{(d+2k-2)}
\gamma^a \xi_{[\mu,\nu]}{}^{a(k-1),b(m)} \nonumber \\
 && \qquad + \frac{2}{(d+2m-2)} \gamma^b 
\xi_{[\mu}{}^{a(k),b(m-1)}{}_{\nu]} \nonumber \\
 && \qquad - \frac{4}{(d+2k-2)(d+2m-2)}
\gamma^a \xi_{[\mu}{}^{a(k-1)b,b(m-1)}{}_{\nu]} ] \label{gaugef6}
\end{eqnarray}
we obtain:
\begin{eqnarray}
\delta_1 {\cal L}_1 &=& - i (-1)^{k+m} 
\frac{8 a_{k,m}{}^2}{5(d+2k-2)(d+2m-4)} \epthree \nonumber \\
&& [ [(d+2k-1)(d+2m-4) + k - m + 2]
\bar{\Psi}_{\mu\nu}{}^{d(k),e(m)} \Gamma^{abc} 
\xi_\alpha{}^{d(k),e(m)} \nonumber \\
&& - 3k [(d+2k-1)(d+2m-4) + 2k - 2m + 4]
\bar{\Psi}_{\mu\nu}{}^{ad(k-1),e(m)} \gamma^b 
\xi_\alpha{}^{cd(k-1),e(m)} \nonumber \\
&& - 3m (d+2k-1)(d+2m-4) \bar{\Psi}_{\mu\nu}{}^{d(k),ae(m-1)} \gamma^b
\xi_\alpha{}^{d(k),ce(m-1)} ]
\end{eqnarray}
Thus in this case it is also impossible to restore the gauge
invariance
by adjusting the parameter $a_{k,m}$.

\subsection{Special case $Y(l+3/2,l+3/2)$}

This time we introduce a two-form $R_{\mu\nu}{}^{a(l),b(l)}$ with
the free Lagrangian:
\begin{eqnarray}
{\cal L}_0 (R_{\mu\nu}{}^{a(l),b(l)}) &=& - i \epfive [ 
\bar{R}_{\mu\nu}{}^{f(l),h(l)} \Gamma^{abcde} \partial_\alpha 
R_{\beta\gamma}{}^{f(l),h(l)} \nonumber \\
 && \qquad - 20l \bar{R}_{\mu\nu}{}^{f(l),ah(l-1)} \Gamma^{bcd} 
\partial_\alpha R_{\beta\gamma}{}^{f(l),eh(l-1)} \nonumber \\
 && \qquad - 60l^2 \bar{R}_{\mu\nu}{}^{af(l-1),b(h(l-1)} \gamma^c 
\partial_\alpha R_{\beta\gamma}{}^{df(l-1),eh(l-1)} ] \label{lagf7}
\end{eqnarray}
which is invariant under the gauge transformations
\begin{equation}
\delta_0 R_{\mu\nu}{}^{a(l),b(l)} = \partial_{[\mu} 
\xi_{\nu]}{}^{a(l),b(l)} + \eta_{[\mu}{}^{a(l),b(l),}{}_{\nu]}
\label{gaugef7}
\end{equation}
Switching to the $AdS$ covariant derivatives we spoil the invariance
of the Lagrangian:
\begin{eqnarray*}
\delta {\cal L}_0 &=& -10 i (d+2l-3)(d+2l-4) \kappa \epthree  \\
 && [ \bar{R}_{\mu\nu}{}^{d(l),e(l)} \Gamma^{abc} 
\xi_\alpha{}^{d(l),e(l)} - 6l \bar{R}_{\mu\nu}{}^{d(l),ae(l-1)}
\gamma^b \xi_\alpha{}^{d(l),ce(l-1)} ]
\end{eqnarray*}
but in this case the invariance can be restored by the introduction of
the mass-like terms into the Lagrangian
\begin{eqnarray}
{\cal L}_1 &=& a_{l,l} \epfour [ \bar{R}_{\mu\nu}{}^{f(l),h(l)}
\Gamma^{abcd} R_{\alpha\beta}{}^{f(l),h(l)} \nonumber \\
 && \qquad \qquad + 12l \bar{R}_{\mu\nu}{}^{f(l),ah(l-1)} \Gamma^{bc}
R_{\alpha\beta}{}^{f(l),dh(l-1)} \nonumber \\
 && \qquad \qquad - 12l^2 \bar{R}_{\mu\nu}{}^{af(l-1),bh(l-1)}
R_{\alpha\beta}{}^{cf(l-1),dh(l-1)} ] \label{lagf8}
\end{eqnarray}
as well as the corresponding corrections to the gauge transformations:
\begin{equation}
\delta_1 R_{\mu\nu}{}^{a(l),b(l)} = \frac{ia_{l,l}}{5(d-4)} [
\gamma_{[\mu} \xi_{\nu]}{}^{a(l),b(l)} + \frac{2}{(d+2l-4)} ( \gamma^a
\xi_{[\mu,\nu]}{}^{a(l-1),b(l)} + \gamma^b 
\xi_{[\mu}{}^{a(l),b(l-1)}{}_{\nu]}) ] \label{gaugef8}
\end{equation}
provided
$$
a_{l,l}{}^2 = - \frac{25}{4} (d+2l-4)^2 \kappa
$$

\section{Infinite spin fermions}

In this section we present an analogous construction for the
fermionic case. As in the bosonic case we begin with the completely
symmetric spin-tensors, then we give a simple example for the mixed
symmetry spin-tensors based on the long hooks and at last we consider
general case based on the mixed symmetry spin-tensors with two rows.

\subsection{Symmetric case --- $Y(k+3/2,1/2)$, $k \to \infty$}

In the recent paper \cite{Met17} Metsaev has shown that the very same
gauge invariant formalism that was previously used for the description
of massive higher spin fermionic fields (see \cite{Met06} for the
metric like formulation and \cite{Zin08b} for the frame-like one) is
also capable of describing a number of infinite spin cases. In this
subsection we reproduce the results of \cite{Met17} but in the
frame-like formalism. We follow our previous results in
\cite{Zin08b} but without a restriction on the number of field
components. Thus we introduce a set of spinor one-forms
$\Phi_\mu{}^{a(k)}$, $0 \le k < \infty$ as well as spinor zero-form
$\phi$. We begin with the sum of kinetic and mass-like terms for all
these fields:
\begin{equation}
{\cal L} = \sum_{k=0}^\infty [ {\cal L}_0(\Phi_\mu{}^{a(k)}) +
{\cal L}_2(\Phi_\mu{}^{a(k)}) ] + i e^\mu{}_a \bar{\phi} \gamma^a 
D_\mu \phi + c_0 \bar{\phi} \phi
\end{equation}
where Lagrangians ${\cal L}_0(\Phi_\mu{}^{a(k)})$ and 
${\cal L}_2(\Phi_\mu{}^{a(k)})$ were defined in (\ref{lagf1}) and
(\ref{lagf2}), with the initial set of gauge transformations defined
in (\ref{gaugef1}) and (\ref{gaugef2}).

Now we add a set of cross-terms gluing all these fields together (see
Figure 4):
\begin{eqnarray}
{\cal L}_1 &=& \sum_{k=1}^\infty (-1)^{k+1} ie_k \eptwo 
( \bar{\Phi}_\mu{}^{ac(k-1)} \gamma^b \Psi_{\nu,c(k-1)} -
\bar{\Phi}_{\mu,c(k-1)} \gamma^a \Phi_\nu{}^{bc(k-1)} ) \nonumber \\
 && + ie_0 e^\mu{}_a ( \bar{\Phi} \gamma^a \phi - \bar{\phi} \gamma^a
\Phi)
\end{eqnarray}
as well as the corresponding corrections to the gauge transformations:
\begin{eqnarray}
\delta_1 \Phi_\mu{}^{a(k)} &=& \frac{e_{k+1}}{6(k+1)} 
\zeta_\mu{}^{a(k)} + \frac{e_k}{6k(d+k-2)} ( e_\mu{}^a \zeta^{a(k-1)}
- Tr ) \nonumber \\
\delta_1 \phi &=& e_0 \zeta
\end{eqnarray}

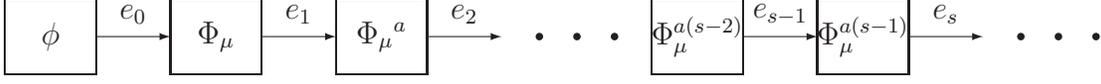
\begin{figure}
\begin{center}
\begin{picture}(150,12)

\put(1,1){\framebox(12,10)[]{$\phi$}}
\put(13,6){\vector(1,0){10}}
\put(15,6){\makebox(6,6)[]{$e_0$}}

\put(23,1){\framebox(12,10)[]{$\Phi_\mu$}}
\put(35,6){\vector(1,0){10}}
\put(37,6){\makebox(6,6)[]{$e_1$}}

\put(45,1){\framebox(12,10)[]{$\Phi_\mu{}^a$}}
\put(57,6){\vector(1,0){10}}
\put(59,6){\makebox(6,6)[]{$e_2$}}

\multiput(72,6)(5,0){3}{\circle*{1}}

\put(87,1){\framebox(12,10)[]{$\Phi_\mu^{a(s-2)}$}}
\put(99,6){\vector(1,0){10}}
\put(101,6){\makebox(6,6)[]{$e_{s-1}$}}

\put(109,1){\framebox(12,10)[]{$\Phi_\mu^{a(s-1)}$}}
\put(121,6){\vector(1,0){10}}
\put(123,6){\makebox(6,6)[]{$e_s$}}

\multiput(136,6)(5,0){3}{\circle*{1}}

\end{picture}
\end{center}
\caption{Completely symmetric fermionic case}
\end{figure}

Then we require that the whole Lagrangian be gauge invariant.
This gives us a number of recurrent relations on the parameters:
$$
b_{k+1} = \frac{(d+2k)}{(d+2k+2)} b_k, \qquad k \ge 0, \qquad 
b_0 = - \frac{3(d-2)}{d} c_0
$$
\begin{eqnarray*}
\frac{(d+2k-2)}{6(k+1)(d+2k)} e_{k+1}{}^2 &=& \frac{e_k{}^2}{6k}
- \frac{2(d+2k-1)}{3(d+2k-2)}b_k{}^2 - \frac{3}{2}(d+2k-1)(d+2k-2)
\kappa \\
\frac{(d-2)}{6d} e_1{}^2 &=& e_0{}^2 - \frac{2(d-1)}{3(d-2)} b_0{}^2
- \frac{3}{2}(d-1)(d-2)\kappa
\end{eqnarray*}

As in the bosonic case  the general solution depends on the two
arbitrary parameters. For the description of massive fermionic 
spin-$(s+1/2)$ field we obtain:
\begin{eqnarray}
b_k{}^2 &=& \frac{9(d+2s-2)^2}{(d+2k)^2} [m^2 - \frac{1}{4} (d+2s-4)^2
\kappa] \nonumber \\
e_k{}^2 &=& \frac{36k(s-k)(d+s+k-2)}{(d+2k-2)} [ m^2 - (s-k-1)
(d+s+k-3) \kappa ] \\
e_0{}^2 &=& \frac{6s(d+s-2)}{(d-2)} [ m^2 - (s-1)(d+s-3)\kappa]
\nonumber
\end{eqnarray}
Let us briefly recall the main properties of this theory. The massless
limit is possible in flat and $AdS$ spaces only. In flat space
such limit results in the sum of massless fields with spins 
$s+1/2,s-1/2, \dots, 1/2$, while in $AdS$ space we get massless
spin-$(s+1/2)$ and massive spin-$(s-1/2)$ fields. In de Sitter
space we have an unitary forbidden region $m^2 <
\frac{1}{4}(d+2s-4)^2\kappa$. Inside this region lives a number of
non-unitary partially massless limits that appears each time then one
of the $e_k = 0$. 

Let us turn now to the infinite spin solution. Here we choose $c_0$
and $e_0$ as our main parameters. We obtain:
\begin{equation}
b_k = \frac{(d-2)c_0}{(d+2k)}
\end{equation}
\begin{equation}
e_k{}^2 = \frac{6k(d+2k-2)}{(d-2)} e_0{}^2 - 
\frac{36k^2(d+k-2)}{(d+2k-2)} c_0{}^2 - 9k^2(d+k-2)(d+2k-2) \kappa
\end{equation}
It is convenient to introduce auxiliary variables $x_k = k(d+k-2)$.
Then we get:
\begin{equation}
e_k{}^2 \sim - 4\kappa x_k{}^2 + [ \frac{8e_0{}^2}{3(d-2)} - 4c_0 -
(d-2)^2\kappa] x_k + \frac{2(d-2)e_0{}^2}{3} \label{cond4}
\end{equation}
The spectrum of the infinite spin solutions is essentially the same
as in the bosonic case. There are no such solutions in de
Sitter space. In flat Minkowski space for the $2e_0{}^2 =
3(d-2)c_0 > 0$ (that again corresponds to $\mu_0 = 0$ in the Metsaev
paper \cite{Met17}) we obtain the massless unitary infinite spin
model, while for $2e_0{}^2 > 3(d-2)c_0 > 0$ --- unitary tahyonic one.
Besides, for $e_0{}^2 < 0$ there exists a discrete spectrum of
solutions which appear each time when we adjust the parameters so that
some $e_s = 0$ and all $e_k{}^2 > 0$, $k > s$:
\begin{eqnarray}
c_0{}^2 &=& \frac{(d+2s-2)^2}{6(d-2)s(d+s-2)}e_0{}^2, \nonumber \\
e_0{}^2 &>& \frac{6s(s+1)(d+s-2)(d+s-1)}{(s-2)} \kappa, \quad s > 0 \\
c_0 &=& 0, \qquad e_0{}^2 > 0, \qquad s = 0 \nonumber
\end{eqnarray}

In Anti de Sitter space we also have two general types of
solution. First of all, there exists the whole range of the parameters
(see Appendix A.5) such that all $e_k{}^2 > 0$ and all our fields
remain to be connected. Besides, there are discrete solutions where
unitary part of the theory contains the fields $\Phi_\mu{}^{a(k)}$
with $k > s$ only.

\subsection{Mixed symmetry example $Y(k+3/2,3/2)$ $k \to \infty$}

In this section we present an example for mixed symmetry fermionic
case based on the long hooks. We follow our previous results on
the massive finite component case \cite{Zin09b}, but without a
restriction on the number of components. Thus we introduce the spinor
two-forms $\Psi_{\mu\nu}{}^{a(k)}$ and spinor one-forms 
$\Phi_\mu{}^{a(k)}$, $0 \le k < \infty$. We begin with the sum of
kinetic and mass-like terms:
\begin{equation}
{\cal L} = \sum_{k=0}^\infty [ {\cal L}_0(\Psi_{\mu\nu}{}^{a(k)}) +
{\cal L}_0(\Phi_\mu{}^{a(k)}) + {\cal L}_2(\Psi_{\mu\nu}{}^{a(k)}) +
{\cal L}_2(\Phi_\mu{}^{a(k)}) ]
\end{equation}
which were defined in (\ref{lagf3}), (\ref{lagf1}), (\ref{lagf4}) and
(\ref{lagf2}) correspondingly. Also we use initial gauge
transformations that were defined in (\ref{gaugef1}), (\ref{gaugef2}),
(\ref{gaugef3}) and ({\ref{gaugef4}).

Let us consider possible cross-terms (see Figure 5).\\
$Y(k+5/2,3/2) \Leftrightarrow Y(k+3/2,3/2)$. In this case the
cross-terms have the form:
\begin{equation}
-i(-1)^k {\cal L}_{11} = c_k \epfour [ 
\bar{\Psi}_{\mu\nu}{}^{ae(k)} \Gamma^{bcd} \Psi_{\alpha\beta}{}^{e(k)}
- \bar{\Psi}_{\mu\nu}{}^{e(k)} \Gamma^{abc} 
\Psi_{\alpha\beta}{}^{de(k)} ]
\end{equation}
while the corresponding corrections to the gauge transformations look
like:
\begin{equation}
\delta_{11} \Psi_{\mu\nu}{}^{a(k)} = \frac{c_k}{10(k+2)} 
\xi_{[\mu,\nu]}{}^{a(k)} - \frac{c_{k-1}}{10k(d+k-3)} ( e_{[\mu}{}^a
\xi_{\nu]}{}^{a(k-1)} - Tr)
\end{equation}

\begin{figure}[htb]
\begin{center}
\begin{picture}(147,32)

\put(1,21){\framebox(12,10)[]{$\Psi_{\mu\nu}$}}
\put(6,21){\vector(0,-1){10}}
\put(6,13){\makebox(6,6)[]{$d_0$}}
\put(1,1){\framebox(12,10)[]{$\Phi_\mu$}}

\put(13,26){\vector(1,0){10}}
\put(15,26){\makebox(6,6)[]{$c_0$}}
\put(13,6){\vector(1,0){10}}
\put(15,6){\makebox(6,6)[]{$e_0$}}

\put(23,21){\framebox(12,10)[]{$\Psi_{\mu\nu}^a$}}
\put(28,21){\vector(0,-1){10}}
\put(28,13){\makebox(6,6)[]{$d_1$}}
\put(23,1){\framebox(12,10)[]{$\Phi_\mu^a$}}

\put(35,26){\vector(1,0){10}}
\put(37,26){\makebox(6,6)[]{$c_1$}}
\put(35,6){\vector(1,0){10}}
\put(37,6){\makebox(6,6)[]{$e_1$}}

\put(45,21){\framebox(12,10)[]{$\Psi_{\mu\nu}^{a(2)}$}}
\put(51,21){\vector(0,-1){10}}
\put(51,13){\makebox(6,6)[]{$d_2$}}
\put(45,1){\framebox(12,10)[]{$\Phi_\mu^{a(2)}$}}

\put(57,26){\vector(1,0){10}}
\put(59,26){\makebox(6,6)[]{$c_2$}}
\put(57,6){\vector(1,0){10}}
\put(59,6){\makebox(6,6)[]{$e_2$}}

\multiput(71,16)(5,0){3}{\circle*{1}}

\put(87,21){\framebox(12,10)[]{$\Psi_{\mu\nu}^{a(s-1)}$}}
\put(92,21){\vector(0,-1){10}}
\put(94,13){\makebox(6,6)[]{$d_{s-1}$}}
\put(87,1){\framebox(12,10)[]{$\Phi_\mu^{a(s-1)}$}}

\put(99,26){\vector(1,0){10}}
\put(101,26){\makebox(6,6)[]{$c_{s-1}$}}
\put(99,6){\vector(1,0){10}}
\put(101,6){\makebox(6,6)[]{$e_{s-1}$}}

\put(109,21){\framebox(12,10)[]{$\Psi_{\mu\nu}^{a(s)}$}}
\put(114,21){\vector(0,-1){10}}
\put(114,13){\makebox(6,6)[]{$d_{s}$}}
\put(109,1){\framebox(12,10)[]{$\Phi_\mu^{a(s)}$}}

\put(121,26){\vector(1,0){10}}
\put(123,26){\makebox(6,6)[]{$c_s$}}
\put(121,6){\vector(1,0){10}}
\put(123,6){\makebox(6,6)[]{$e_s$}}

\multiput(136,16)(5,0){3}{\circle*{1}}

\end{picture}
\end{center}
\caption{Simplest mixed symmetry fermionic example}
\end{figure}
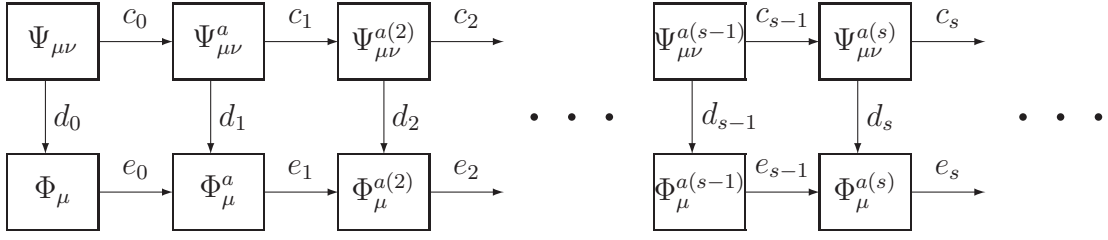

$Y(k+3/2,3/2) \Leftrightarrow Y(k+3/2,1/2)$. Here we introduce the
following terms:
\begin{eqnarray}
-i(-1)^k {\cal L}_{12} &=& d_k \epthree [ 
\bar{\Psi}_{\mu\nu}{}^{d(k)} \Gamma^{abc} \Phi_\alpha{}^{d(k)} - 6k
\bar{\Psi}_{\mu\nu}{}^{ad(k-1)} \gamma^b \Phi_\alpha{}^{cd(k-1)} ]
\nonumber \\
 && + d_k \epthree [ \bar{\Phi}_\mu{}^{d(k)} \Gamma^{abc}
\Psi_{\nu\alpha}{}^{d(k)} - 6k \bar{\Phi}_\mu{}^{ad(k-1)} \gamma^b
\Psi_{\nu\alpha}{}^{cd(k-1)} ]
\end{eqnarray}
as well as the corresponding corrections to the gauge transformations:
\begin{eqnarray}
\delta_{12} \Psi_{\mu\nu}{}^{a(k)} &=& - 
\frac{(k+1)d_k}{10(k+2)(d-3)(d-4)}
[ \Gamma_{\mu\nu} \zeta^{a(k)} - \frac{1}{2(k+1)(d+k-3)} \cdot
\nonumber \\
 && ((d-3)(d-4)+2k+2) e_{[\mu}{}^{(a} \zeta_{\nu]}{}^{a(k-1)} -
(d+2k-1) \gamma^{(a} \gamma_{[\mu} \zeta_{\nu]}{}^{a(k-1)}) ] \\
\delta \Phi_\mu{}^{a(k)} &=& 2d_k \xi_\mu{}^{a(k)} \nonumber
\end{eqnarray}

$Y(k+5/2,1/2) \Leftrightarrow Y(k+3/2,1/2)$. The last type of the
cross-terms look like:
\begin{equation}
-i(-1)^k {\cal L}_{13} = e_k \eptwo [ \bar{\Phi}_\mu{}^{ac(k)}
\gamma^b \Phi_\nu{}^{c(k)} - \bar{\Phi}_\mu{}^{c(k)} \gamma^a 
\Phi_\nu{}^{bc(k)} ]
\end{equation}
with the corresponding corrections having the form:
\begin{equation}
\delta_{13} \Phi_\mu{}^{a(k)} = \frac{e_k}{6(k+1)} \zeta_\mu{}^{a(k)}
+ \frac{e_{k-1}}{6k(d+k-2)} ( e_\mu{}^a \zeta^{a(k-1)} - Tr )
\end{equation}

Now we require that the total Lagrangian be invariant under the
resulting gauge transformations. This gives a number of recurrent
relations on the parameters:
$$
(d+2k+2)a_{k+1} = (d+2k)a_k
$$
$$
(k+2)(d+l-1)d_{k+1}{}^2 = (k+1)(d+k-2)d_k{}^2
$$
$$
b_k = - \frac{3(d-2)}{5(d-4)}a_k, \qquad
e_k{}^2 = \frac{9(k+1)(d+k-1)}{25(k+2)(d+k-2)}c_k{}^2
$$
$$
\frac{2(d-3)(d-4)}{5(k+1)(k+2)(d+k-2)(d+2k)} c_k{}^2
= \frac{4(k+1)}{(k+2)}d_k{}^2 - \frac{8(d-3)}{5(d-4)}a_k{}^2 -
10(d-3)(d-4)\kappa
$$

For the massive finite spin case where some $c_s = 0$ (and hence $e_s
= 0$) we obtain the following solution (here $m$ is not the mass, but
just a convenient parameter):
\begin{eqnarray*}
d_k{}^2 &=& \frac{(s+1)(d+s-2)}{(k+1)(d+k-2)}m^2 \\
c_k{}^2 &=& \frac{10(k+1)(d+s+k)(s-k)}{(d+2k)} [
\frac{(d-4)(s+1)}{(d-3)(s+2)}m^2 + 10(k+2)(d+k-2)\kappa] \\
a_k{}^2 &=& \frac{(d+2s)^2(d-4)}{4(d+2k)^2(d-3)(s+2)}
[ 10(s+1)(s+d-2)m^2 - 25(d-3)(d-4)(s+2)\kappa]
\end{eqnarray*}
The massless limit (that requires $c_{s-1} \to 0$ and $d_s \to 0$
simultaneously) is possible in flat Minkowski space only. In de Sitter
space there is an unitary forbidden region
$$
m^2 < \frac{5(d-4)(d-3)(s+2)}{2(s+1)(d+s-2)} \kappa 
$$
Inside this region there exists the only partial massless limit 
$d_s \to 0$ (and hence all $d_k \to 0$) when the whole system
decomposes into two disconnected ones containing fields from the upper
and lower rows in the Figure 5. In Anti de Sitter space we also
find an unitary forbidden region
$$
m^2 < - \frac{10(d-3)(s+2)(d+s-3)}{(d-4)} \kappa
$$
At the boundary lives the only unitary partially massless limit where
$c_{s-1} = 0$ (and hence $e_{s-1} = 0$). In this limit two main fields
$\Psi_{\mu\nu}{}^{a(s)}$ and $\Phi_\mu{}^{a(s)}$ decouple. Inside the
unitary forbidden region there is a number of partially massless
limits that appear each time when one of the $c_l = 0$ (and hence $e_l
= 0$).

For the infinite spin solutions we choose the parameters $c_0$ and
$d_0$ as our main ones (see Figure 5). Then we obtain:
\begin{eqnarray*}
a_k{}^2 &=& \frac{50}{8(d+2k)^2} [ - 4(d-4)^2 C + 4d^2 D - d^2(d-4)^2
\kappa] \\
c_k{}^2 &=& \frac{200(k+1)}{2(d+2k)} [ (k+2)(d+k-2) C - k(d+k) D \\
 && \qquad \qquad - k(k+2)(d+k)(d+k-2)\kappa ] \\
d_k{}^2 &=& \frac{(d-2)}{(k+1)(d+k-2)} d_0{}^2
\end{eqnarray*}
where
$$
C = \frac{dc_0{}^2}{200(d-2)}, \qquad D = \frac{(d-4)d_0{}^2}{20(d-3)}
$$
If we introduce the auxiliary variables $x_k = k(d+k)$ and function
\begin{equation}
F(x_k) = - \kappa x_k{}^2 + [ C - D - d\kappa ] x_k + 2dC
\label{cond5}
\end{equation}
then we will find that
\begin{eqnarray*}
c_k{}^2 &\sim& F(x_k), \qquad k \ge 0 \\
d_k{}^2 &\sim& F(x_l), \qquad l = - 2 \\
a_k{}^2 &\sim& F(x_l), \qquad l = - \frac{d}{2}
\end{eqnarray*} 
In flat Minkowski space we obtain the unitary infinite spin
solution where all our fields enter for the quite restricted range of
the parameters:
$$
C \ge D \ge \frac{(d-4)^2}{d^2} C
$$

In Anti de Sitter space there is a region of parameters (see
Appendix A.6) when all our fields remain to be connected. At the
boundary of this region there is a limit when all $d_k = 0$ 
so that Figure 5 splits vertically into two parts containing fields
$\Psi_{\mu\nu}$ and $\Phi_\mu$ respectively. Besides, there is a
discrete set of solutions where some $c_s = 0$ (and hence $e_s = 0$)
while all $c_k{}^2 > 0$ for $k > s$ so that Figure 5 splits
horizontally with the unitary part containing fields 
$\Psi_{\mu\nu}{}^{a(k)}$ and $\Phi_\mu{}^{a(k)}$ with $k > s$ only.
Moreover, these solutions admit a limit where all $d_k = 0$ and
Figure 5 splits into four disconnected parts.

\subsection{General case $Y(k+3/2,l+3/2)$, $k \to \infty$}

We need the following set of fields: $\Psi_{\mu\nu}{}^{a(k),b(m)}$, $0
\le m \le l$, $l \le k \le \infty$, and $\Phi_\mu{}^{a(k)}$, $l \le k
\le \infty$. The sum of their kinetic and mass-like terms is:
\begin{equation}
{\cal L} = \sum_{k=l}^\infty [ \sum_{m=0}^l 
({\cal L}_0(\Psi_{\mu\nu}{}^{a(k),b(m)}) + 
{\cal L}_2(\Psi_{\mu\nu}{}^{a(k),b(m)})) + 
{\cal L}_0(\Phi_\mu{}^{a(k)}) + {\cal L}_2(\Phi_\mu{}^{a(k)}) ]
\end{equation}
where all these Lagrangians as well as corresponding gauge
transformations were defined in the previous section.

Now we introduce all possible cross-terms\footnote{The meaning of the
coefficients is essentially the same as in the bosonic case, see
Figure 3.}.\\
$Y(k+3/2,m+3/2 \Leftrightarrow Y(k+1/2,m+3/2)$:
\begin{eqnarray}
{\cal L}_{11} &=& i \sum_{k=l+1}^\infty \sum_{m=1}^l (-1)^{k+m}
c_{k,m} \epfour [ \bar{\Psi}_{\mu\nu}{}^{ae(k-1),f(m)}
\Gamma^{bcd} \Psi_{\alpha\beta}{}^{e(k-1),f(m)} \nonumber \\
 && \qquad \qquad - 6m \bar{\Psi}_{\mu\nu}{}^{ae(k-1),bf(m-1)}
\gamma^c \Psi_{\alpha\beta}{}^{e(k-1),df(m-1)} \nonumber \\
 && \qquad \qquad - \bar{\Psi}_{\mu\nu}{}^{e(k-1),f(m)} \Gamma^{abc}
\Psi_{\alpha\beta}{}^{de(k-1),f(m)} \nonumber \\
 && \qquad \qquad - 6m \bar{\Psi}_{\mu\nu}{}^{e(k-1),af(m-1)}
\Gamma^{bcd} \Psi_{\alpha\beta}{}^{ce(k-1),df(m-1)} ]
\end{eqnarray}
Corrections to the gauge transformations:
\begin{eqnarray}
\delta_{11} \Psi_{\mu\nu}{}^{a(k),b(m)} &=& - 
\frac{c_{k+1,m}}{10(k-m+2)(k+2)} [(k-n+2) \xi_{[\mu,\nu]}{}^{a(k)b(m)}
+ \xi_{[\mu}{}^{a(k)b,b(m-1)}{}_{\nu]}] \nonumber \\
 && - \frac{c_{k,m}}{10k(d+k-3)} [ e_{[\mu}{}^a 
\xi_{\nu]}{}^{a(k-1),b(m)} + \dots  ]
\end{eqnarray}

$Y(k+3/2,m+3/2) \leftrightarrow Y(k+3/2,m+1/2)$:
\begin{eqnarray}
{\cal L}_{12} &=& i \sum_{k=l}^\infty \sum_{m=1}^l (-1)^{k+m}
d_{k,m} \epfour [ \bar{\Psi}_{\mu\nu}{}^{e(k),af(m-1)}
\Gamma^{bcd} \Psi_{\alpha\beta}{}^{e(k),f(m-1)} \nonumber \\
 && \qquad \qquad + 6k \bar{\Psi}_{\mu\nu}{}^{ae(k-1),bf(m-1)}
\gamma^c \Psi_{\alpha\beta}{}^{de(k-1),f(m-1)} \nonumber \\
 && \qquad \qquad - \bar{\Psi}_{\mu\nu}{}^{e(k),f(m-1)} \Gamma^{abc}
\Psi_{\alpha\beta}{}^{e(k),df(m-1)} \nonumber \\
 && \qquad \qquad + 6k \bar{\Psi}_{\mu\nu}{}^{ae(k-1),f(m-1)}
\gamma^b \Psi_{\alpha\beta}{}^{ce(k-1),df(m-1)} ] 
\end{eqnarray}
Corrections:
\begin{eqnarray}
\delta_{12} \Psi_{\mu\nu}{}^{a(k),b(m)} &=& - 
\frac{d_{k,m}}{10m(k-m+2)(d+m-4)} [ (k-m+1)
\xi_{[\mu}{}^{a(k),b(m-1)} e_{\nu]}{}^b \nonumber \\
&& - e_{[\nu}{}^a \xi_{\mu]}{}^{a(k-1)b,b(m-1)} + \dots ] 
 - \frac{d_{k,m+1}}{10(m+1)} \xi_{[\mu}{}^{a(k),b(m-1)}{}_{\nu]}
\end{eqnarray}

There are three special cases 
$Y(k+5/2,3/2) \Leftrightarrow Y(k+3/2,3/2)$,
$Y(k+3/2,3/2) \Leftrightarrow Y(k+3/2,1/2)$,
$Y(k+5/2,1/2) \Leftrightarrow Y(k+3/2,1/2)$ that have been considered
in the previous subsection.

Now we require that the whole Lagrangian be gauge invariant. First of
all, this gives a number of important relations on the parameters:
\begin{eqnarray*}
a_{k,m} &=& \frac{(d+2l)(d+2l-2)}{(d+2k)(d+2m-2)} a_{l,l} \\
c_{k,m}{}^2 &=& \frac{(k-l)(d+k+l-2)}{(k-m)(d+k+m-2)} c_{k,l}{}^2 \\
d_{k,m}{}^2 &=& \frac{(l-m+1)(d+l+m-2)}{(k-m+1)(d+k+m-2)} d_{l,m}{}^2
\\
e_k{}^2 &=& \frac{9(k-l)(d+k+l-2)}{25(k+1)(d+k-3)} c_{k,l}{}^2
\end{eqnarray*}
Thus all the parameters are determined by the $a_{l,l}$, $c_{k,l}$, $k
\ge l+1$ and $d_{l,m}$, $0 \le m \le l$. Further, for these parameters
we obtain:
\begin{eqnarray*}
c_{k,l}{}^2 &=& \frac{k(d+2k-2)}{(k+1)(d+2k)} c_{k+1,l}{}^2
 + \frac{4k(d+2k-1)}{(d+2k-2)} a_{k,l}{}^2 \\
 && + 25k(d+2k-1)(2+2k-2)\kappa \\
d_{l,m}{}^2 &=& \frac{m(l-m)((d+2m-8)(d+l+m-1) + 6)}
{(m+1)(l-m+1)(d+2m-2)(d+l+m-2)} d_{l,m+1}{}^2 \\
 && + \frac{m(d^2+3dl+dm-7d+2l^2+2lm-8l-6m+12)}{(l+1)(d+2l)(d+l+m-2)} 
c_{l+1,m}{}^2 \\
 && + \frac{8m(d+2l-1)(d+2m-4)+l-m+2)}{(d+2l-2)(d+2m-4)} a_{l,m}{}^2 
\\
 && + 50m((d+2l-3)(d+2m-4) + (l-m)(2l-2m+3))\kappa
\end{eqnarray*}

For the massive mixed symmetry spin-tensor $Y(s+3/2,l+3/2)$ we obtain
the following solution:
\begin{eqnarray*}
a_{l,l}{}^2 &=& \frac{(d+2s)^2}{4(d+2l)^2}[m^2 - 25(d+2l-4)^2 \kappa ]
\\
c_{k,l}{}^2 &=& \frac{k(s-k+1)(d+s+k-1)}{(d+2k-2)} [ m^2
+ 100(k-l+1)(d+k+l-3)\kappa ] \\
d_{l,n}{}^2 &=& \frac{n(s-n+2)(d+s+n-2)(d+l+n-3)}{(d+2n-4)(d+l+n-2)}
[ m^2 - 100(l-n)(d+l+n-4)\kappa ]
\end{eqnarray*}
Once again the massless limit is possible in flat Minkowski space
only. In de Sitter space we obtain an unitary forbidden region
$$
m^2 < 25(d+2l-4)^2\kappa
$$
Inside this region we find a number of non-unitary partially massless
limits where some $d_{l,k} = 0$. In Anti de Sitter space there is
also an unitary forbidden region
$$
m^2 < - 100(s-l+1)(d+s+l-3)\kappa
$$
At the boundary of this region lives the only unitary partially
massless case where $c_{s,l} = 0$, while inside this region there
exists a number of non-unitary partially massless models where some
$c_{k,l} = 0$.

For the infinite spin solutions we choose $c_{l+1,l}$ and $d_{l,0}$ as
our main parameters. Then we obtain:
\begin{eqnarray*}
a_{l,l}{}^2 &=& - \frac{25(d-4)^2}{(d+2l)^2} C + 25 D
- \frac{25(d-4)^2}{4}\kappa \\
c_{k,l}{}^2 &=& \frac{100k}{(d+2k-2)} [ (k+1)(d+k-3) C 
- (k-l-1)(d+k+l-1) D \\
 && \qquad - (k+1)(d+k-3)(k-l-1)(d+k+l-1)\kappa ] \\
d_{l,k}{}^2 &=& \frac{100k(d+l+k-3)}{(d+2k-4)(d+l+k-2)}
[ k(d+k-4) C + (l-k+2)(d+l+k-2) D \\
 && \qquad + k(l-k+2)(d+k-4)(d+l+k-2)\kappa ]
\end{eqnarray*}
where
$$
C = \frac{(d+2l)c_{l+1,l}{}^2}{100(l+1)(l+2)(d+l-2)}, \qquad
D = \frac{(d-4)d_{l,0}{}^2}{10(l+2)(d+l-3)}
$$
Introducing variables $x_k = k(d+k-4)$ and the function
\begin{equation}
F(x_k) = - \kappa x_k{}^2 + [C - D + (l+2)(d+l-2)\kappa] x_k -
(l+2)(d+l-2)D \label{cond6}
\end{equation}
we find:
\begin{eqnarray*}
c_{k-1,l}{}^2 &\sim& F(x_k), \qquad l+2 \le k < \infty \\
d_{l,k}{}^2 &\sim& F(x_k), \qquad 0 \le k \le l \\
a_{l,l}{}^2 &\sim& F(x_k), \qquad k = - \frac{(d-4)}{2}
\end{eqnarray*}
For flat Minkowski space there exists an unitary region of
parameters:
$$
\frac{(d+2l)^2}{(d-4)^2}D \ge C \ge D
$$
where all our fields remain to be connected.

In Anti de Sitter space there exists a range of the parameters
(see appendix A.7) where unitary theory contains all our fields.
Besides, there exist two sets of discrete solutions. The first one
corresponds to the situations when some $c_{s,l} = 0$ (and hence all
$c_{s,m} = 0$ for $0 \le m \le l$) so that the whole set of fields
decomposes into finite and infinite parts, the infinite part being
unitary. The second type of solutions corresponds to the cases when
some $d_{l,s} = 0$ (and hence all $d_{k,s} = 0$ for $ k \ge l$) so
that the whole set of fields splits into two infinite parts, one of
them being unitary. Moreover, there are special points where
simultaneously $c_{s,l} = 0$ and $d_{l,l} = 0$ when our system
decomposes into four disconnected parts.

\appendix

\section{On the unitary solutions in $AdS$ space}

\subsection{Method of resolving the unitarity conditions}

In every case, the unitarity condition of the Lagrangian is given by a
system of linear inequalities, where each of them has the form
\mbox{$(A_i)^2_{(k_i)}\ge0$}, where $A_i$ stands for the letter of the
coefficient and $k_i$ stands for its ordinal number. The unitarity
condition of the partially massless limit is given by a system of
linear inequalities and linear equations:
\mbox{$(A_i)^2_{(k_i)}\ge0$,$(B_j)^2_{(k_j)}=0$}. Then, it turns out
that the sign of any coefficient is determined by a sign of a square
trinomial
\begin{equation}
P(x) = - \kappa x^2 + (B\kappa + C)x + D
\end{equation}
at some $x$, wherein $B$ depends only on the dimension of the
space-time, while $C$ and $D$ are two independent linear combinations
of squares of two coefficients of Lagrangian. Let $A_1$ and $A_2$ be
the coefficients. Hence, the unitarity condition is a system of linear
inequalities of two variables. The solution of such system is a convex
set, on whose border one or two inequalities become equalities.
Let us sort the conditions by $x$: $P(x_i)\sim 0$, where
"$\sim$"\mbox{ }is either "$\ge$" \mbox{ } or "$=$". Then, we will
frequently use two following assertions:

\begin{quote}
1. If a square trinomial $P(x)$ with positive leading coefficient is
such that conditions $P(x_0)=0,P(x_1)\ge0$ and $x_0<x_1$ hold, then
for any $x>x_1$ the inequality $P(x)>0$ holds. The same is true for
$x<x_1<x_0$
\end{quote}
\begin{quote}
2. If a square trinomial $P(x)$ with positive leading coefficient is
such that condition $P(x_1)=P(x_2)=0$ holds and $x_1<x_2$ , then
$P(x)<0$ for any $x\in(x_1,x_2)$ and $P(x)>0$ for any $x<x_1\vee
x>x_2)$.
\end{quote}

Then in case of $\kappa<0$ the second assertion implies that all the
points determined by expressions $P(x_i)=0,P(x_{i+1})=0$, and only
they are the vertices of the region of unitarity. Hence, the boundary
of the region of unitarity is a union of segments of lines $P(x_i)=0$
whose ends are points $P(x_i)=0,P(x_{i-1})=0$ and
$P(x_i)=0,P(x_{i+1})=0$. The first and the last conditions, if they
exist, give rays instead of segments. The region of unitarity can be
defined as a union of regions given by inequalities
$A_1\in[(A_1)_k,(A_1)_{k+1}],A_2\in[f_k(A_1),g_k(A_1)]$, where
$(A_1)_k$ are coordinates of vertices of the region of unitarity, in
ascending order, and $f_k$, $g_k$ are either $\pm \infty$ or linear
functions obtained by resolving equations  $P(x_i)= 0$ with the
respect to $A_2$. The convexity of the region of unitarity implies
that $g_k(A_1) \ge f_k(A_1)$ for $A_1\in[(A_1)_k,(A_1)_{k+1}]$.

As to partially massless limits, each of them is given by one or two
equations and a system of inequalities. In case of one  equation
$P(x_i)=0$, the assertion 1 implies that all of the inequalities,
except $P(x_{i-1})=0$ and $P(x_{i+1})=0$ if they exist, can be removed
without altering the solution. Hence, the unitary partially massless
limit is either a segment or a ray, whose ends are points where
$P(x_{i-1})=0$ and $P(x_{i+1})=0$ intersect $P(x_i)=0$. Similarly to
the case of complete Lagrangian, the area can be expressed either in
form $A_2=aA_1+b,A_1\in [(A_1)_1,(A_1)_2]$ or in form
$A_1=(A_1)_0,A_2\in [(A_2)_1,(A_2)_2]$. In case of two equalities the
assertion 2 implies that the solution is non-empty if and only if one
equality is next to the other in the list. In that case, the solution
is a unique point.

An example of different areas is given in Figure 1. Here, grey area is
the region of unitarity, given by inequalities $P(x_i)=0,i \in
\overline{1,4}$, red segment is the partially massless limit given by
conditions
\mbox{$P(x_1)\ge 0,P(x_2)= 0,P(x_3)\ge 0,P(x_4)\ge 0$}, green ray the
partially massless limit given by conditions
\mbox{$P(x_3)= 0$},\mbox{$P(x_4)\ge 0$}, and the blue dot is a double
partially massless limit given by conditions
\mbox{$P(x_1)= 0$},\mbox{$P(x_3)= 0,P(x_4)\ge 0$}.

\setlength{\unitlength}{0.8pt}
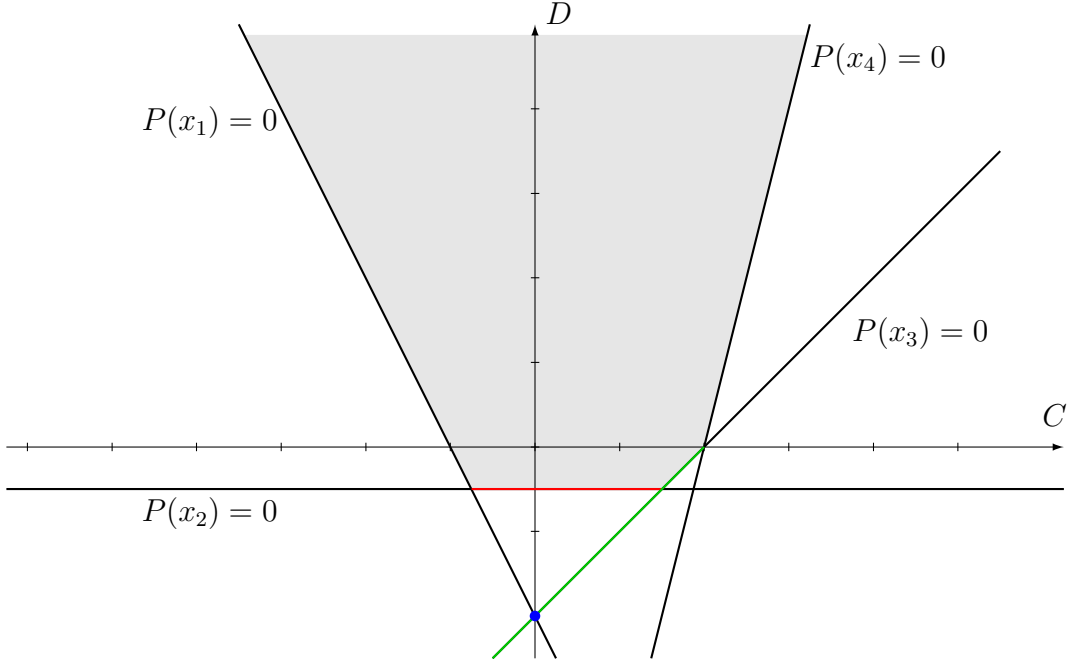
\begin{figure}[ht]
\begin{center}
\begin{picture}(500,300)
\color[rgb]{0,0,0}
\thinlines
\color[rgb]{0.9,0.9,0.9}
\put(0,0){\polygon*(113,295)(220,80)(310,80)(330,100)(379,295)}
\color[rgb]{0.0,0.0,0.0}
\put(250,0){\vector(0,1){300}}
\put(0,100){\vector(1,0){500}}
\color[rgb]{0.0,0.0,0.0}
\put(255,300){$D$}
\put(490,110){$C$}
\put(64,250){$P(x_1)=0$}
\put(64,65){$P(x_2)=0$}
\put(400,150){$P(x_3)=0$}
\put(380,280){$P(x_4)=0$}
\multiput(10,98)(40,0){12}{\line(0,1){4}}
\multiput(248,20)(0,40){7}{\line(1,0){4}}
\thicklines
\color[rgb]{0.0,0.0,0.0}
\put(305,0){\line(1,4){75}}
\put(260,0){\line(-1,2){150}}
\put(0,80){\line(1,0){500}}
\put(230,0){\line(1,1){240}}
\color[rgb]{1.0,0.0,0.0}
\put(220,80){\line(1,0){90}}
\color[rgb]{0.0,0.8,0.0}
\put(230,0){\line(1,1){100}}
\color[rgb]{0.0,0.0,1.0}
\put(250,20){\circle*{5}}
\color[rgb]{0,0,0}
\end{picture}
\end{center}
\caption{Example of different unitarity regions in $AdS$ space}
\end{figure}

In the flat case, all the inequalities become homogeneous. Hence, the
unitarity region is an angle with the vertex at the origin,i.e. point
$A_1=A_2=0$. The list of inequalities contains an infinite number of
conditions for increasing and unbounded sequence of $x_i$. The region
of unitarity thus is defined by inequalities $Cx_1+D\ge0,C\ge0$. In
contrast to the $AdS$ case, the trivial case $P(x_i)=0$ is possible in
flat space. It corresponds to values $A_1=A_2=0$ and defines totally
massless limit. Any non-trivial partially massless limit is given by
one equation $P(x_i)=0$ and an infinite list of inequalities. Such
limit is possible if and only if the equation is the first element of
the list of conditions. The solution in that case is  $P(x_i)=0,C>0$.

\subsection {Completely symmetric tensor $Y(k+1,0)$}

In this case unitarity requires that all $e_k{}^2 \ge 0$, 
$k\in\overline{0,+\infty}$. Applying the general method to the
function (\ref{cond1}) we find that the region of unitarity (such that
all our fields enter) is determined by the following set of
conditions:
\begin{eqnarray}
e_0^2 &\in& \bigg[\frac{2s(s+1)(s+d-2)(s+d-3)|\kappa|}{(d-2)},
\frac{2(s+1)(s+2)(s+d-2)(s+d-11)|\kappa|}{(d-2)}\bigg], \nonumber \\
c_0 &\ge&
\frac{s(d+s-1)\big[e_0^2+2(s+1)(d+s-2)|\kappa|\big]}{(s+1)(d+s-2)},
\quad s\in\overline{0,+\infty}
\end{eqnarray}
Besides, there exists a discrete set of unitary solutions that appear
each time when some $e_s = 0$, $s > 0$, while all $e_k{}^2 > 0$, 
$k > s$:
\begin{eqnarray}
c_0 &=& \frac{(s-1)(d+s-2)\big[e_0^2+2s(s+d-3)|\kappa|\big]}{s(s+d-3)}
, \nonumber \\
e_0^2 &\ge& \frac{2s(s+1)(s+d-3)(s+d-2)|\kappa|}{(d-2)}
\end{eqnarray}
as well as a special case
\begin{equation}
c_0 \ge 0, \qquad e_0^2 = 0.
\end{equation}

\subsection {Mixed symmetry tensor $Y(k+1,1)$}

In this case unitarity requires that all $e_k{}^2 \ge 0$, $d_k{}^2 \ge
0$, $k\in\overline{0,+\infty}$. The general analysis applied to the
function (\ref{cond2}) shows that the unitarity region is determined
be the following set of conditions:
\begin{eqnarray}
C &\in&
\bigg[\frac{k(k-1)(k+d-2)(k+d-1)|\kappa|}{2(d-3)},\frac{k(k+1)(k+d-1)(k+d)|\kappa|}{2(d-3)}\bigg], \nonumber \\
D &\in&
\bigg[0,\frac{(k+2)(k+d-3)(C+k(k+d-1)|\kappa|)}{k(k+d-1)}\bigg], \quad
 k\in\overline{1,+\infty}
\end{eqnarray}
At the boundary of this region corresponding to $D = 0$ and $C > 0$
all $d_k{}^2 = 0$ so that the whole system decomposes into two
subsystems with the fields $\Psi_{\mu\nu}$ and $\Phi_\mu$
respectively. There exists also a set of discrete solutions when some
$c_s = 0$ (and hence $e_s = 0$), while all $c_k{}^2 > 0$, $k > s$:
\begin{eqnarray}
C &=& \frac{(k-1)(k+d-2)(D-(k+1)(k+d-4)|\kappa|)}{(k+1)(k+d-4)},
\nonumber \\
D &\in& \bigg(0,\frac{(k+1)(k+2)(k+d-4)(k+d-3)|\kappa|}{2(d-3)}\bigg)
\end{eqnarray}
In this case the unitary part of the theory contains fields
$\Psi_{\mu\nu}{}^{a(k)}$ and $\Phi_\mu{}^{a(k)}$ with $k > s$. Note
that in this case it is also possible to take a limit $D \to 0$ so
that the whole system decomposes into four independent ones where the
subsystem  with the fields $\Psi_{\mu\nu}{}^{a(k)}$, 
$s \le k < \infty$ provides one more non-trivial example of the
unitary infinite spin theory.

\subsection {General bosonic case $Y(k+1,l+1)$}

In this case unitarity requires that all $c_{k,l}{}^2 \ge 0$, $k > l$
and $d_{l,k}{}^2 \ge 0$, $0 \le k \le l$. All these parameters are
given by the values of the one and the same function (\ref{cond3}), so
that our general method applicable here as well. We find that unitary
region of parameters is determined by the following set of conditions:
\begin{eqnarray*}
C &\in& \bigg[\frac{k(k+1)(d+2l+k-1)(d+2l+k)|\kappa|}{d+2l-3},
\frac{(k+1)(k+2)(d+2l+k)(d+2l+k+1)|\kappa|}{d+2l-3}\bigg], \\
D &\le& \frac{(k+3)(d+2l+k-2)(2(k+1)(d+2l+k)|\kappa|+C)}
{(k+1)(d+2l+k)}, \qquad k\in\overline{0,+\infty}, \\ \\
C &\in& \bigg[\frac{3}{5}(d+2l-4)|\kappa|\bigg], \qquad
 D \ge 0 \\
C &\in&
\bigg[\frac{(k+2)(k+3)(d+2l-k-4)(d+2l-k-3)|\kappa|}{d+2l-3}, \\
 && \frac{(k+3)(k+4)(d+2l-k-5)(d+2l-k-4)|\kappa|}{d+2l-3}\bigg], \\
D &\ge& \frac{(k+1)(d+2l-k-6)(C-2(k+3)(d+2l-k-4)|\kappa|)}
{(k+3)(d+2l-k-4)}, \quad k\in\overline{0,l-1}, \\
C &\in&
\bigg[\frac{(l+1)(l+2)(d+l-3)(d+l-2)|\kappa|}{d+2l-3},+\infty\bigg),
\\
D &\ge& \frac{l(l+d-5)(C-2(l+2)(d+l-3)|\kappa|)}{(l+2)(l+d-3)}
\end{eqnarray*}
Besides, there are two types of discrete solutions. The first one
corresponds to the cases where some $c_{c,l} = 0$ (and hence all
$c_{s,m} = 0$, $0 \le m \le l$):
\begin{eqnarray}
C &\in& \bigg[-2(s-l-2)(s+l+d-3)|\kappa|, \nonumber \\
 &&
\frac{(s-l-2)((s-l-1)(s+l+d-3)(s+l+d-2))|\kappa|}{(d+2l-3)}\bigg],\\
D &=& \frac{( s-l)( s+l+d-5)(C-( s-l-2)( s+l+d-3) |\kappa|) }
{( s-l-2)(s+l+d-3) } \nonumber
\end{eqnarray}
so that  Figure 3 splits horizontally. The second one corresponds to
the cases where some $d_{l,m} = 0$ (and hence all $d_{k,m} = 0$, $l
\le k < \infty$):
\begin{eqnarray}
C &\ge& \frac{(l+2-s)((l+1-s)(s+l+d-3)(s+l+d-2))|\kappa|}{(d+2l-3)},
\nonumber \\
D &=& \frac{(l-s)(s+l+d-5)(C-(l+2-s)( s+l+d-3)|\kappa|)}
{(l+2-s)(s+l+d-3)}, \qquad s < l \\
C &\ge& 0, \qquad D = 0, \qquad s = l \nonumber
\end{eqnarray}
when Figure 3 splits vertically. Moreover, these two sets have common
points when simultaneously some $c_{s,l} = 0$ and $d_{l,l} = 0$:
\begin{equation}
D = 0, \qquad C = - 2(s-l-1)(s+l+d-2)|\kappa|
\end{equation}
so that Figure 3 splits into four disconnected parts.

\subsection {Completely symmetric spin-tensor $Y(k+3/2,0)$}

In this case unitarity requires that all $e_k{}^2 \ge 0$ for $k \ge
0$. Using our general method for the function (\ref{cond4}) we find
the the unitarity region is determined by the following set of
parameters:
\begin{eqnarray}
e_0^2 &\in& \bigg[\frac{6s(s+1)(s+d-1)(s+d-2)|\kappa|}{(d-2)},
\frac{6(s+1)(s+2)(s+d)(s+d-1)|\kappa|}{(d-2)}\bigg], \nonumber \\
c_0 &\in& [0,
\frac{(d+2s-2)^2\big[2e_0^2+3(d-2)s(d+s-2)|\kappa|\big]}
{12(d-2)s(d+s-2)}], \quad s\in\overline{0,+\infty},
\end{eqnarray}
Besides, there exists a set of discrete solutions where some $e_s = 0$
while all $e_k{}^2 > 0$ for $k > s$:
\begin{eqnarray}
c_0 &=&
\frac{(d+2s-2)^2\big[2e_0^2+3(d-2)s(d+s-2)|\kappa|\big]}
{12(d-2)s(d+s-2)}, \nonumber \\
e_0^2 &<&
\frac{(d+2s-2)^2\big[2e_0^2+3(d-2)s(d+s-2)|\kappa|\big]}
{12(d-2)s(d+s-2)}, \quad s > 0 \\
e_0 &=& 0, \qquad c_0^2 < \frac{d^2|\kappa|}{4}, \quad s = 0 \nonumber
\end{eqnarray}
with  unitary part containing fields $\Phi_\mu{}^{a(k)}$ with $k > s$
only.

\subsection {Mixed symmetry spin-tensor $Y(k+3/2,3/2)$}

In this case unitarity requires that all $c_k{}^2 \ge 0$, $d_k{}^2 \ge
0$ and $a_k{}^2 \ge 0$ for $k\in\overline{0,+\infty}$. All these
parameters are related with the values of one and the same function
(\ref{cond5}) so that our general method is valid. We obtain for the
unitary region the following set of conditions:
\begin{eqnarray*}
C &\in& \bigg[\frac{k(k+1)(k+d)(k+d+1)|\kappa|}{2(d-2)},
\frac{(k+1)(k+2)(k+d+1)(k+d+2)|\kappa|}{2(d-2)}\bigg], \\
D &\le& \frac{(k+3)(k+d-1)(C+(k+1)(k+d+1)|\kappa|)}{(k+1)(k+d+1)},
\quad k\in\overline{0,+\infty}, \\ \\
C &\in& \bigg[0,\frac{|\kappa d^2|}{4}\bigg], \qquad
 D \ge 0 \\
C &\ge& \frac{|\kappa d^2|}{4}, \qquad
D \ge \frac{(d-4)^2(4C-\kappa d^2)}{4d^2}
\end{eqnarray*}
At the boundary of this region there is a limit when all $d_k = 0$,
namely:
\begin{equation}
D = 0, \qquad \frac{d^2|\kappa|}{4}\ge C > 0
\end{equation}
so that Figure 5 splits vertically into two parts containing fields
$\Psi_{\mu\nu}$ and $\Phi_\mu$ respectively. Besides, there is a
discrete set of solutions where some $c_s = 0$ (and hence $e_s = 0$)
while all $c_k{}^2 > 0$ for $k > s$:
\begin{eqnarray}
D &=& \frac{(s+2)(s+d-2)(C+s(s+d)|\kappa|)}{s(s+d)}, \nonumber \\
C &\in& \bigg[-s(s+d)|\kappa|,
\frac{s(s+1)(s+d)(s+d+1)|\kappa|}{2(d-2)}\bigg], \quad s > 0 \\
C &=& 0, \qquad D \in [0,-3(d-1)|\kappa|], \quad s = 0 \nonumber
\end{eqnarray}
so that Figure 5 splits horizontally with the unitary part containing
fields $\Psi_{\mu\nu}{}^{a(k)}$ and $\Phi_\mu{}^{a(k)}$ with $k > s$
only. Moreover, these solutions admit a limit
\begin{equation}
D = 0, \qquad C = - s(s+d)|\kappa|
\end{equation}
where all $d_k = 0$ and Figure 5 splits into four disconnected parts.

\subsection {General fermionic case $Y(k+3/2,l+3/2)$}

In this case unitarity requires that $c_{k,l}{}^2 \ge 0$, $d_{l,m}{}^2
\ge 0$, $a_{l,l}{}^2 \ge 0$, $k\in\overline{l+1,+\infty}$,
$n\in\overline{0,l}$. All of these parameters are related with the
different values of the one and the same function (\ref{cond6}). Using
our general method we find that the unitarity region is determined by
the following set of conditions:
\begin{eqnarray*}
C &\in&
\bigg[\frac{s(s+1)(2l+s+d-1)(2l+s+d)|\kappa|}{(l+2)(l+d-2)}, \\
&& \frac{(s+1)(s+2)(2l+s+d+1)(2l+s+d)|\kappa|}{(l+2)(l+d-2)}\bigg], \\
D &\le& \frac{(s+l+2)(d+s+l-2)(|\kappa|s(d+2l+s)+C)}{s(d+2l+s)}, \quad
s\in\overline{0,+\infty}, \\ \\
C &\in& \bigg[0,\frac{6|\kappa|(d+2l-3)(d+2l-2)}{(l+2)(l+d-2)}\bigg],
\\
D &\ge& \frac{(l-1)(l+d-5)(3(2l+d-3)|\kappa|-C)}{3(2l+d-3)}, \\ \\
C &\in& \bigg[\frac{(s+2)(s+3)(d+2l-s-2)(d+2l-s-3)|\kappa|} 
{(l+2)(l+d-2)}, \\
 && \frac{(s+3)(s+4)(d+2l-s-3)(d+2l-s-4)|\kappa|}
{(l+2)(l+d-2)}\bigg], \\
D &\ge& \frac{(l-s-1)(d+l-s-5)(|\kappa|(s+3)(d+2l-s-2)-C)}
{(s+3)(d+2l-s-2)}, \quad s\in\overline{0,l-2},  \\ \\
C &\in& \bigg[(l+1)(l+d-1)|\kappa|,\frac{(d+2l)^2|\kappa|}{4}\bigg),
\qquad D \ge 0,\\ \\
C &\in& \bigg(\frac{(d+2l)^2|\kappa|}{4},+\infty\bigg),\qquad
D \ge \frac{(d-4)^2\big[4C+|\kappa|(d+2l)^2\big]}{4(d+2l)^2}
\end{eqnarray*}
Besides, there exist two sets of discrete solutions. The first one
corresponds to the situation when some $c_{s-1,l} = 0$ (and hence all
$c_{s-1,m} = 0$ for $0 \le m \le l$):
\begin{eqnarray}
C &\in& \bigg[-\frac{(s-l-2)(s+l+d-2)(2l+d-2)|\kappa|}{(l+2)(d+l-2)},
\nonumber \\
 &&
\frac{(s-l-2)(s-l-1)(s+l+d-2)(s+l+d-1)|\kappa|}{(l+2)(d+l-2)}\bigg],
\nonumber \\
D &=& \frac{(s-l)(s+d-4)(C-(s-l-2)(s+l+d-2)|\kappa|) }
{(s-l-2)(s+l+d-2)}, \quad s > l+2 \\
C &=& 0, \qquad D \in \big[|\kappa|l(l+d-4),|\kappa|(l+3)(l+d-1)\big],
\quad s = l+2 \nonumber
\end{eqnarray}
where the whole set of fields decomposes into the finite and infinite
parts, the infinite part being unitary. The second type of solutions
corresponds to the cases when some $d_{l,s} = 0$ (and hence all
$d_{k,s} = 0$ for $k \ge l$):
\begin{eqnarray}
D &=& \frac{s(s+d-4)((l+2-s)(s+l+d-2)|\kappa|-C)}{(l+2-s)(s+l+d-2)}
\nonumber \\
C &\in& \bigg[\frac{(l+2-s)(s+l+d-2)(d+2l)^2|\kappa|}{4(l+2)(l+d-2)},
\nonumber \\
&&
\frac{|\kappa|(l+2-s)(l+1-s)(s+l+d-2)(s+l+d-1)}{(l+2)(l+d-2)}\bigg],
\quad s < l \\
D &=& \frac{l(l+d-4)(2(2l+d-2)|\kappa|-C)}{2(2l+d-2)}, \nonumber \\
C &\in& \bigg[0,\frac{|\kappa|(d+2l-2)(d+2l)^2}{2(l+2)(l+d-2)}\bigg],
\quad s = l \nonumber
\end{eqnarray}
so that the whole set of fields splits into two infinite parts, one of
them being unitary. Moreover, there are special points where
simultaneously $c_{s,l} = 0$ and $d_{l,l} = 0$:
\begin{eqnarray}
D &=& \frac{|\kappa|ls(l+d-4)(s+d-4)}{(l+2)(l+d-2)}, \nonumber \\
C &=& -\frac{2|\kappa|(2l+d-2)(s-l-2)(s+l+d-2)}{(l+2)(l+d-2)}
\end{eqnarray}
when our system decomposes into four disconnected parts.

\end{document}